\documentclass{emulateapj}
\usepackage{multirow}
\usepackage{natbib}

\begin{document}

\title{Simulating the Cooling Flow of Cool-Core Clusters}
\author{Yuan Li and Greg L. Bryan}
\affil{Department of Astronomy, Columbia University, Pupin Physics Laboratories, New York, NY 10027}

\begin{abstract}
We carry out high-resolution adaptive mesh refinement simulations of a cool core cluster, resolving the flow from Mpc scales down to pc scales.  We do not (yet) include any AGN heating, focusing instead on cooling in order to understand how gas gets to the supermassive black hole (SMBH) at the center of the cluster.  We find that, as the gas cools, the cluster develops a very flat temperature profile, undergoing a cooling catastrophe only in the central 10-100 pc of the cluster.  Outside of this region, the flow is smooth, with no local cooling instabilities, and naturally produces very little low-temperature gas (below a few keV), in agreement with observations.  The gas cooling in the center of the cluster rapidly forms a thin accretion disk.  The amount of cold gas produced at the very center grows rapidly until a reasonable estimate of the resulting AGN heating rate (assuming even a moderate accretion efficiency) would overwhelm cooling.  We argue that this naturally produces a thermostat which links the cooling of gas out to 100 kpc with the cold gas accretion in the central 100 pc, potentially closing the loop between cooling and heating.   Isotropic heat conduction does not affect the result significantly, but we show that including the potential well of the brightest cluster galaxy is necessary to obtain the correct result.  Also, we found that the outcome is sensitive to resolution, requiring very high mass resolution to correctly reproduce the small transition radius. 
\end{abstract}


\section{Introduction}

The majority of baryonic matter in galaxy clusters resides in the form of virialized hot gas that emits in the X-ray band \citep[e.g.][]{ Lin}. Observations of the X-ray surface brightness, along with imaging spectroscopy, allow the measurement of the density and temperature of the intracluster medium (ICM). Many, if not most, relaxed clusters possess a cool core (CC), where the temperature drops by a factor of 2 to 3 compared to the cluster outskirts and the radiative cooling time is much shorter than the Hubble time \citep[e.g.][]{Cavagnolo08, Sanderson09_2}.  In a steady state, a ``cooling flow'' of $100s$ M$_{\odot}/$yr would develop in a typical CC cluster \citep[see review by][]{Fabian94}. However, XMM grating spectra have revealed a dearth of gas at temperatures lower than about one-third of the cluster virial temperature, despite the short cooling time of the gas \citep{Tam01, Peterson03, Sanders08}. In addition, the observed star-formation rate \citep[e.g.][]{Cardiel} is typically a factor of $\sim 10-100$ times lower than the mass deposition rate predicted by classic cooling flow models. This discrepancy implies that there is some process that heats the gas, preventing it from cooling. Many mechanisms have been proposed, including thermal conduction \citep[e.g.][]{ZN03}, active galactic nuclei (AGN) feedback \citep[e.g.][]{Bohringer93, McNamara05, BM06}, MHD wave-mediated cosmic rays combined with conduction \citep[e.g.][]{Loewenstein, GuoOh08}, and turbulence combined with conduction \citep[e.g.][]{DC05}.

AGN feedback is considered to be the most plausible heating source for a number of reasons \citep{McNamara07}.  First, there is an observed correlation between the presence of cool cores and signs of AGN activity \citep[e.g.][]{DF06}.  In addition, in nearby CC clusters, X-ray cavities or bubbles are observed in the core \citep[e.g.][]{Fabian00} and are thought to be inflated by the interactions between powerful jets and the surrounding gas.  Finally, the estimated energy injected into the surrounding gas by the bubbles can balance cooling in most of these systems \citep[e.g.][]{Birzan04}. However, how the gas feeds the AGN, i.e. whether the gas accretion is dominated by the ``hot'' mode described in the Bondi-Hoyle accretion model \citep{Bondi,Allen06} or by the ``cool'' mode assuming that the hot gas first fragments within the sonic radius and falls onto the black hole as cold clouds \citep[e.g.][]{BS89}, and how AGN deposits the energy to the ICM (whether kinetic or thermal feedback dominates) are still open questions -- see \citet{Croton} and references therein. To address these issues, we need to first understand exactly how gas cools and eventually accretes onto the black hole.

Theoretical work has been carried out using linear perturbation theory to investigate the local cooling instability \citep[e.g.][]{Binney, BS89}, deriving the global dynamics of the cooling flow in a steady state \citep[e.g.][]{QN2000}, or simulating the global cooling flow numerically \citep[e.g.][]{CT07, Gaspari10, Guo10}. However, due to its complex and non-linear nature, an accurate picture of the cooling flow cannot be derived from analytical models, and the previous simulations usually suffer from a low resolution with finest grid sizes typically around a kpc.

In this paper, we carry out 3D simulations using the adaptive-mesh hydro code Enzo to investigate the onset of the cooling flow. Our primary objective is to explore the cooling catastrophe in detail and examine how gas cools and flows inwards in the cluster center.  We do not attempt to model the accretion on to the SMBH itself and the resulting feedback, and so our model is not a complete description of thermally-balanced feedback; however, by examining the detailed evolution of a cluster as it undergoes a cooling catastrophe, we can get a better idea of the evolution and production of cold gas in the cluster core.  Key issues that this study aims to address include: (i) confirmation of linear calculations that local cooling instabilities do not grow significantly before a cooling catastrophe occurs, (ii) the location and amount of cold gas produced by the global thermal instability, (iii) the rate of gas accretion on to a central SMBH, (iv) observational signatures during the cooling catastrophe (in particular, the lack of cool gas observed in X-rays). (v) the impact of other processes (such as isotropic thermal conduction and Type Ia SN heating rates) on the cooling catastrophe.

This paper is organized as follows, in Section~\ref{sec:method}, we describe our code, initial conditions and (non-standard) refinement method; in Section~\ref{sec:results}, we present the main results from our standard simulation; in Section~\ref{sec:discussion}, we discuss the related physical and numerical issues, and compare our results with observations and previous work. We conclude in Section~\ref{sec:conclusion}.

% --------------------------------------------

\section{Methodology}
\label{sec:method}

The simulations are performed using the Enzo Adaptive Mesh Refinement (AMR) code, a parallel, Eulerian hydrodynamics scheme. An important advantage of using an AMR code is that it achieves high levels of resolution by only refining areas that need it and thus reduces computational time. The code initially places a uniform root or ``parent'' grid consisting of relatively large cells over the whole simulation box. If further resolution is required, a finer ``child'' grid is placed inside the parent cell and the properties of each of its grid cells are then computed. The child grid then becomes a parent grid at the next step and this process can be repeated until the desired level of resolution is reached. A more detailed description of Enzo can be found in \citet{B97} and \citet{Oshea}, and references therein.

The hydrodynamics algorithm we choose for most of our runs is a three-dimensional version of the Zeus astrophysical code developed by \citet{ZEUS} in Cartesian coordinate system. It is a simple, fast, and robust algorithm that allows large problems to be run at high resolution. For comparison, we also perform test runs using the piecewise parabolic method (PPM) \citep{PPM}.

In the following, we describe the initial setup of the simulations in Section \ref{sec:methodology_initial}, our refinement strategy in Section \ref{sec:methodology_refinement}, and the physics included in the simulations in Section \ref{sec:methodology_physics}.

% --------------------------------------------

\subsection{Initial Conditions}\label{sec:methodology_initial}

\begin{figure}
\begin{center}
\includegraphics[width=0.5\textwidth]{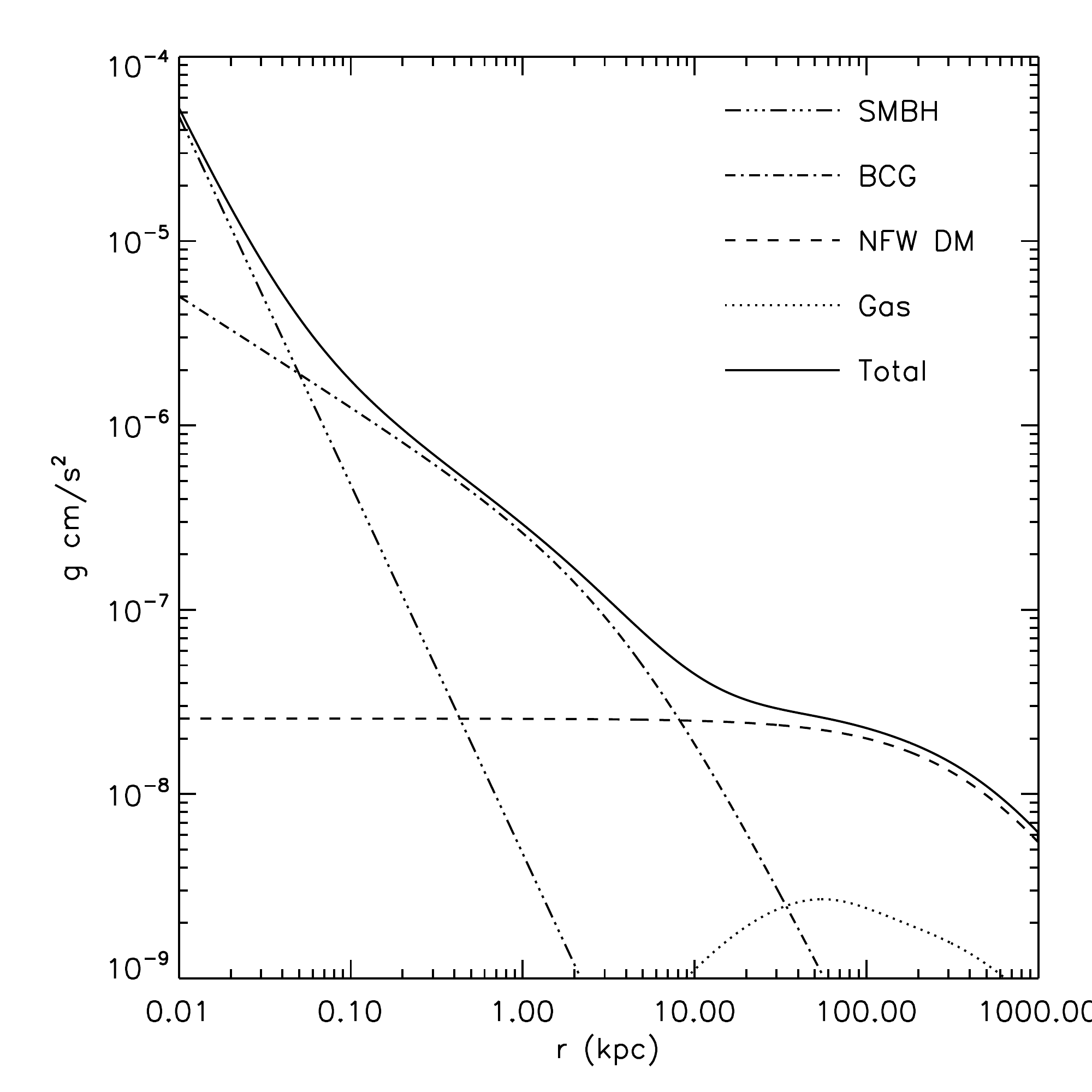}
\caption{The gravitational acceleration for our idealized cluster (solid line) along with the contribution from different components.  All contributions except for the gas are static and do not change during the simulation. 
\label{fig_g}}
\end{center}
\end{figure}

Our simulations start with an isolated idealized galaxy cluster roughly in hydrostatic equilibrium. We adopt the Perseus Cluster, a massive relaxed CC cluster that has been well observed, as our model template. Initially the gas is taken to be spherically symmetric and the azimuthally averaged temperature profile is taken from analytic fits to the observations of the Perseus Cluster \citep{Churazov}:
\begin{equation}
T = 7 \; \frac{1 + (r/71)^3}{2.3 + (r/71)^3} \; \rm{keV}\;,
\end{equation}
where $r$ is the radius to the center of the cluster in kpc.  This temperature profile is flat for $r > 0.5$ Mpc, while Perseus, like many clusters, has a declining profile at large radius \citep[e.g.][]{Vikhlinin06}. Since we are interested in the evolution in the center ($r < 0.2$ Mpc), we do not expect this difference to be significant. The electron density profile \citep{Mathews} is taken to be:
\begin{equation}
n_e(r) = \frac{0.0192}{1 + \left(\frac{r}{18}\right)^3} + \frac{0.046}{\left[1 + \left(\frac{r}{57}\right)^2\right]^{1.8}}
+ \frac{0.0048}{\left[1 + \left(\frac{r}{200}\right)^2\right]^{1.1}} \rm{cm}^{-3},
\end{equation}
The power-law index of the last term is slightly steepened compared to \citet{Mathews} so that at small radii (up to a few hundred kpc) the gas density profile does not significantly deviate from that obtained from the X-ray observation, but at large radii, where the fit is poorly constrained, the density drops as $r^{-2.2}$, which is more consistent with cosmological simulations and fits better with the NFW dark matter profile \citep{NFW}. We compute the initial pressure from the density and temperature assuming the ideal gas law with $\gamma = 5/3$. 

The gravitational potential is the combination of the self-gravity of the ICM (which does not dominate but is included for completeness), and three static components: the dark matter halo, the stellar mass of the brightest cluster galaxy (BCG) and the super-massive black hole (SMBH) in its center. The dark matter halo follows a NFW distribution that matches the observed gas density and temperature of Perseus:
\begin{equation}
\rho^{NFW} (r) = \frac{\rho^{NFW}_0}{(\frac{r}{r_s})(1 + \frac{r}{r_s})^2} \;\;,
\end{equation}
where $\rho^{NFW}_0 = 7.5 \times 10^{14}$ M$_{\odot}$/Mpc$^3$, and $r_s = 0.494$ Mpc is the scale radius. Since the gas density and temperature are only observed up to a few hundred kpc and are uncertain at large radii, the NFW parameters can vary depending on the outer radius we choose when fitting the model. We experimented with slightly different NFW parameters and the results were not affected. 

The stellar mass profile of the BCG is:
\begin{equation}
M_*(r) = 
\frac{r^2}{G} \left[ \left( {r^{0.5975} \over 3.206 \times 10^{-7}}\right)^s
+ \left( {r^{1.849} \over 1.861\times 10^{-6}}\right)^s 
\right]^{-1/s}
\end{equation}
in cgs units with $s = 0.9$ and $r$ in kpc \citep{Mathews}.  The SMBH is treated as a point mass at the very center of the cluster with $M_{\rm SMBH} = 3.4 \times 10^8$ M$_{\odot}$ \citep{BHmass}. 

The self-gravity of the ICM is computed at each step, but it does not have much impact in our major runs because the total mass of the cool gas in the cluster center is small compared to the BCG and SMBH. The mass of the accumulated gas in the center of the core only becomes significant at late time in our low resolution simulations where the cooling flow is stronger (see Section \ref{sec:resolution} for discussion). The gravitational acceleration from different components is plotted in Figure~\ref{fig_g}. As can be seen from that figure, the radius of influence of the SMBH (where its acceleration dominates) occurs at approximately 50 pc.

In order to give the gas some initial angular momentum, we set the cluster initially rotating slowly around the z axis in our simulation. The angular momentum of a galaxy cluster can be characterized by its spin parameter $\lambda$, which was first introduced by \citet{Peebles} and later interpreted also as $\lambda = \omega/\omega_{sup}$ \citep{spin}, where $\omega_{sup}$ is the angular velocity that would provide rotational support to the system. As a simplification, we use the same rotational velocity $v$ at all radii and $\omega \sim v/r_s$. Thus the spin parameter $\lambda$ can be expressed as $\lambda = v / v_c $, where $v_c = \sqrt{\frac{G M_s}{r_s}}$ with $M_s$ being the total mass inside the scale radius $r_s$. Typical values of $\lambda$ for a cluster are around 0.05 \citep[e.g.][]{Bullock}. For our model cluster, we take a constant rotational velocity of $v = 50$ km/s, which is consistent with cluster simulations. 

We also give each cell an initial random velocity that could potentially seed small-scale instabilities. The random velocity obeys a Gaussian distribution with a standard deviation of 200 km/s for each of the three components, a typical value for the turbulent motion of gas in galaxy clusters \citep{RO10}.

To set up the initial cluster configuration, we employ an iterative technique: starting with the root grid, we use the equations above to set the density, temperature and velocities.  We then apply the refinement criteria discussed in the next section to obtain additional levels of refinement, applying the initial conditions to each level and reapplying the initial conditions.  The result is a grid hierarchy which is self-consistently refined, and contains high-resolution initial data.

% --------------------------------------------

\subsection{Refinement Strategy}\label{sec:methodology_refinement}

We refine a cell whenever any of the following three criteria are met:

(i) The gas mass criterion -- a cell is refined if the gas mass in any cell exceeds 0.2 times the gas mass in one cell of the root grid, or more precisely, when:
\begin{equation}
\label{eq:m_cell}
m_{cell} > 0.2 \;\rho_{\rm mean} (\frac{L}{N_{\rm root}})^3 \times 2^{\alpha l} \;\;,
\end{equation}
where $\rho_{mean}$ is the mean density of the universe at the initial time, $L = 16$ Mpc is the co-moving box size, $l$ is the level of refinement, and $N_{\rm root}$ is the number of cells on the root grid in each dimension. We use $N_{\rm root} = 256$ for our standard simulation.  This defines a critical gas mass that is always refined, which we then modify (with the factor $2^{\alpha l}$) depending on the level of refinement.   If $\alpha=0$, the refinement is Lagrangian and the limit for $m_{\rm cell}$ does not change with refinement level $l$; in this case, a given mass clump will be resolved by the same number of cells as it collapses to smaller sizes.   A negative value of $\alpha$ makes the refinement ``super-Lagrangian'' and the maximum value of $m_{\rm cell}$ decreases with $l$. The decrease is faster as $\alpha$ becomes more negative.  In our simulation, we find that choosing a more negative $\alpha$ can better resolve the early evolution of the cooling flow (see Section \ref{sec:resolution} for discussion). We use $\alpha = -1.2$ for our standard run, which reduces the maximum mass of a cell by a factor of $3.8\times 10^{-6}$ going from $l = 0$ to $l = 15$. In the standard run, the level of refinement in the initial conditions is $l = 8$, which gives a maximum cell mass (i.e. roughly the mass resolution in the initial conditions) of about $8\times 10^5$ M$_{\odot}$. 

(ii) The cooling criterion -- we also apply refinement whenever the ratio of cooling time to sound-crossing time over the cell becomes too small (i.e. $t_{\rm cool}/t_{\rm cross}(\Delta x) < \beta$). The isobaric cooling time for an optically-thin plasma is given by 
\begin{equation}
  \label{eqtcool} 
t_{\rm cool} = \frac{\frac{5}{2} n k_b T}{n^2 \Lambda(T)},
\end{equation}
where $k_b$ is the Boltzmann's constant, $n = \rho/\mu m_H$ is the particle number density and $\Lambda(T)$ is the cooling function \citep{CoolingFunction}. The sound crossing time over a cell is $t_{\rm cross}(\Delta x) = \Delta x/c_s$, where $\Delta x$ is the cell size, and $c_s=\sqrt{\gamma P/\rho}$ is the sound speed.  We use this criterion because when the cooling time is shorter than the sound-crossing time, the gas drops out of pressure equilibrium, and the cooling becomes isochoric.  We use the limit $\beta = 6$, which is a somewhat arbitrarily chosen value larger than 1 so that we can fully resolve this transition. We have experimented with higher value of $\beta = 12$ and $\beta = 20$ for this ratio and the results do not change. 

(iii) The Jeans length criterion -- a cell is refined whenever its size is larger than $1/4$ of the Jeans length, to ensure that we properly resolve any gravitational instabilities \citep{Truelove}. 

In our simulations, the baryon mass criterion is important at early stages, well before the cooling catastrophe occurs, while the cooling criterion becomes important later as the gas density grows in the center, and the radiative cooling becomes more significant.  The Jeans criterion is never important for our standard runs.

% --------------------------------------------

\begin{figure*}
\begin{center}
\includegraphics[scale=.4]{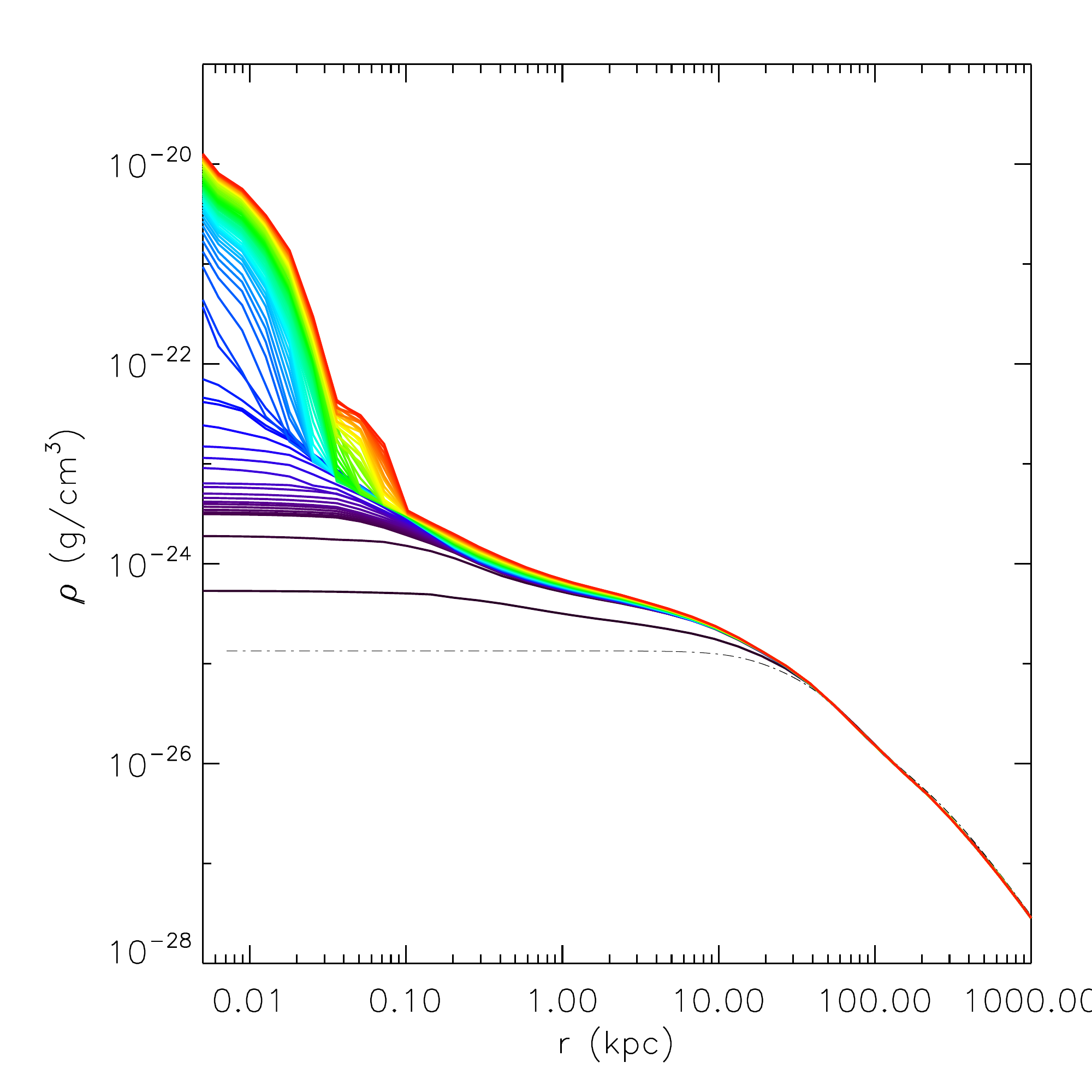}\includegraphics[scale=.4]{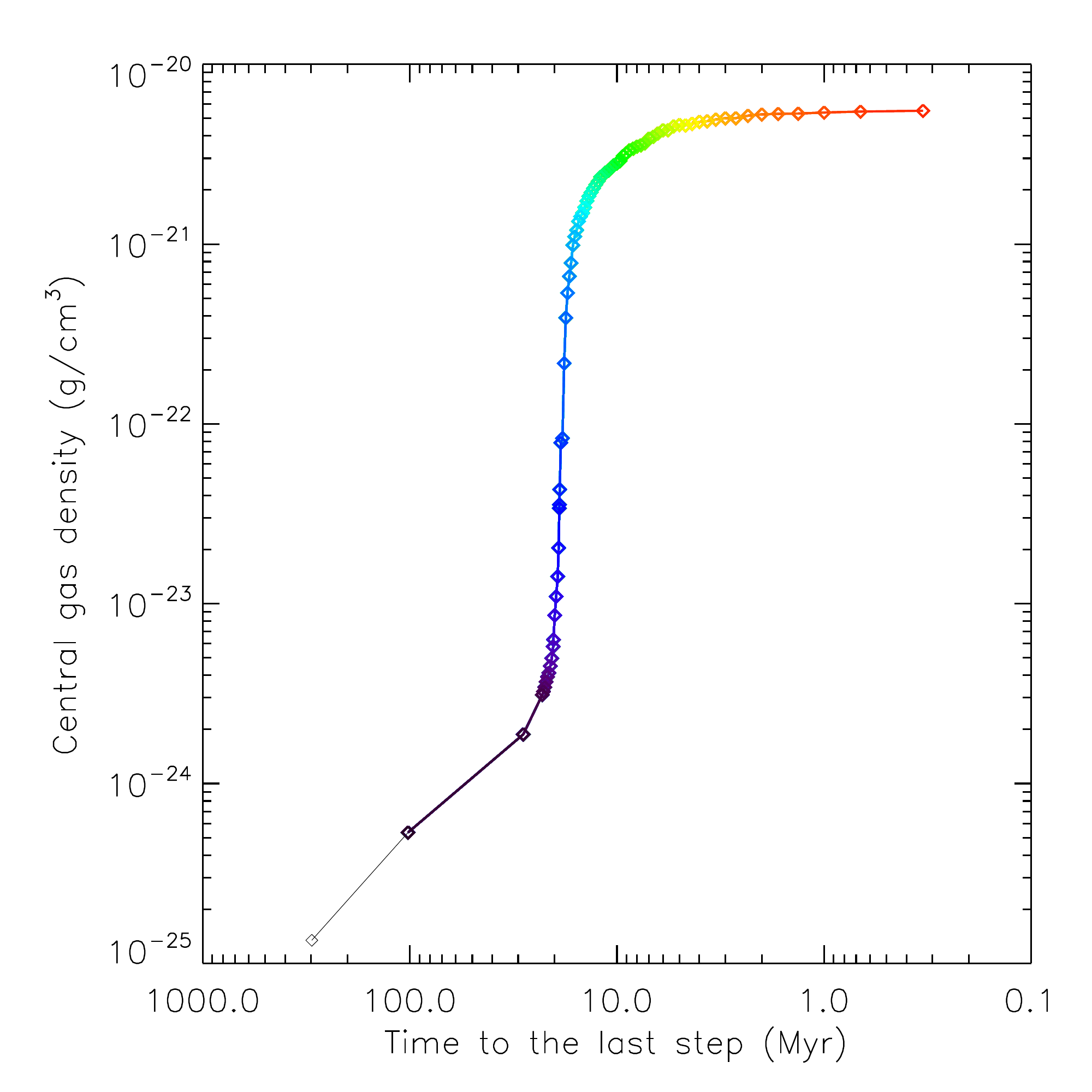}
\includegraphics[scale=.4]{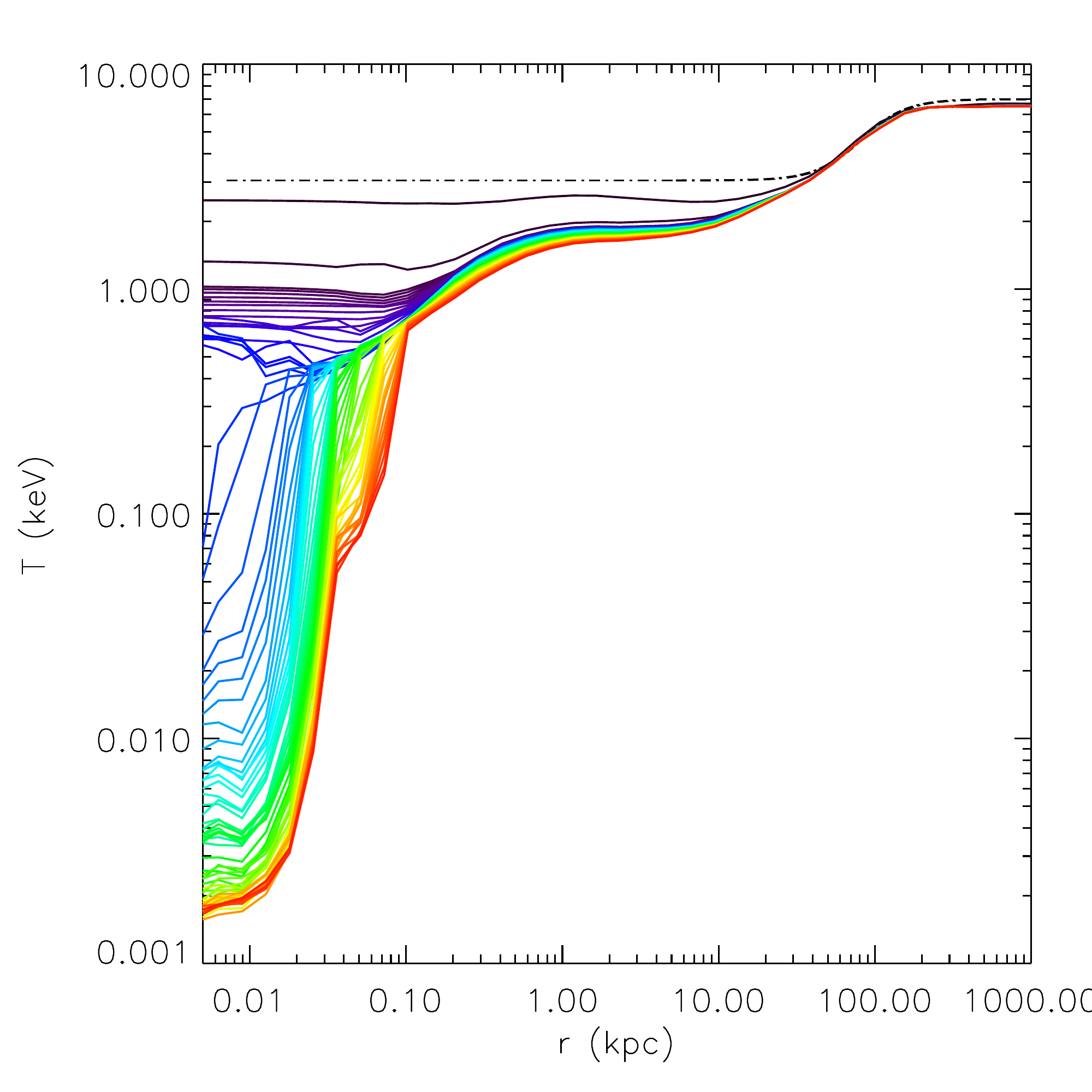}\includegraphics[scale=.4]{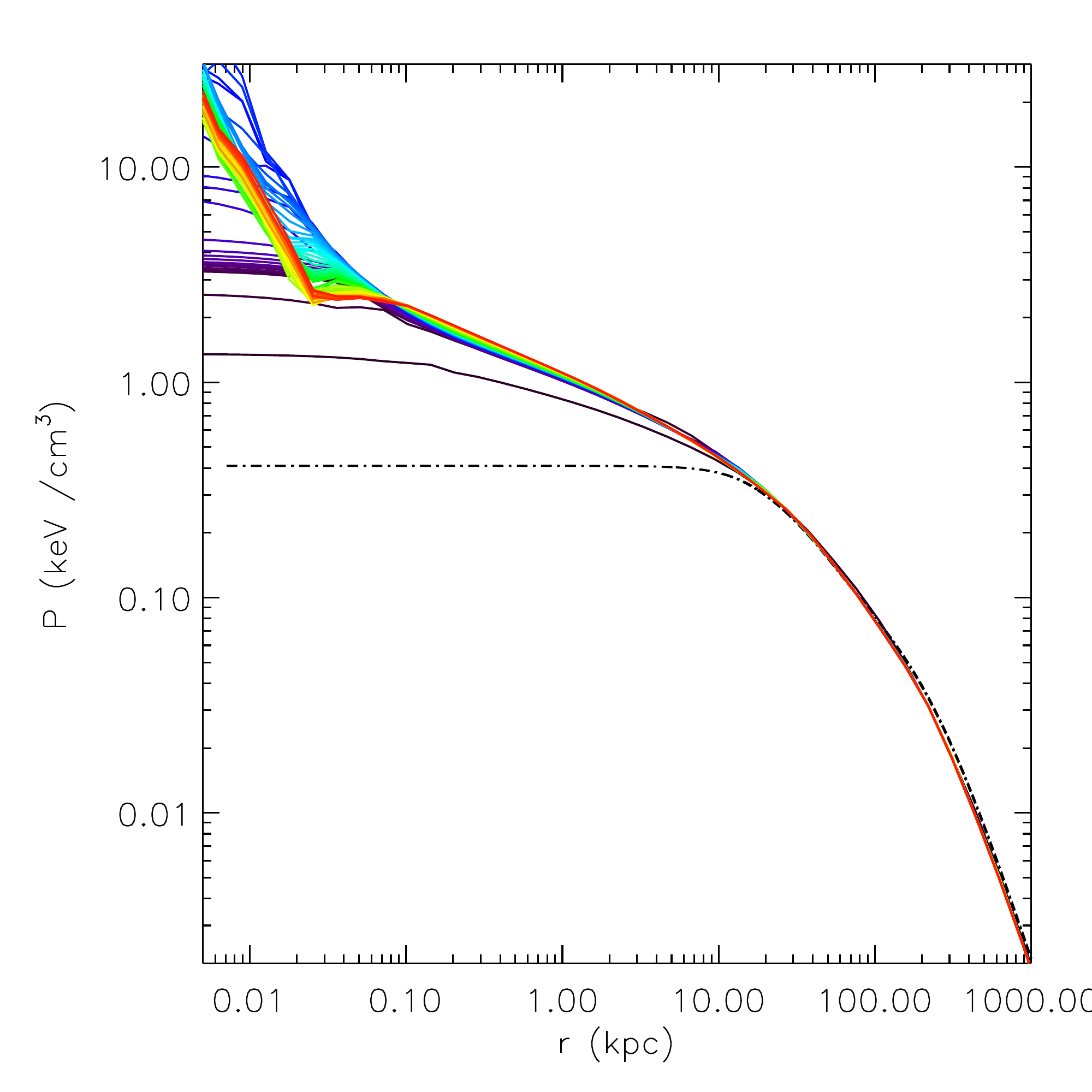}
\caption{The evolution of gas density, temperature and pressure.  In the top row, these are: (left) spherically averaged gas density profiles, and (right) the time evolution of the central gas density.  On the bottom row, we plot (left) the mass weighted temperature profile, and (right) the gas pressure profile. The black dashed lines correspond to the initial conditions.  Each solid line is from one output, with different colors corresponding to the different times on the right panel. \label{fig_rho}}
\end{center}
\end{figure*}

\subsection{The Simulations}\label{sec:methodology_physics}

All simulations in this paper include radiative cooling.  The radiative cooling function we use is adapted from the SPEX package \citep{CoolingFunction}, assuming equilibrium cooling.  Non-equilibrium effects will play a role in cooler gas, but as this only occurs in the very center of our simulations, we do not expect this to play an important role.   We adopt a metallicity of one-half solar for our model cluster \citep{Metallicity}. The cooling is truncated at temperatures lower than $10^4 K$, because we are not interested in following the evolution of gas below this temperature.

Our standard run, which will be the primary simulation analyzed in this paper, has $N_{\rm root} = 256$, with a co-moving box size of $L = 16$ Mpc, and a maximum refinement level of $l_{\rm max} = 15$.  The minimum physical size of a grid cell in this simulation is given by $\Delta x_{min} = L / (N_{\rm root} 2^{l_{\rm max}})$, which is about $2.0$ pc.

To test the dependence of the results on the resolution, we also performed simulations with $N_{root} = 64$ , $128$ and $256$.  We also carry out simulations with the same $N_{root}$ but different values of $\alpha$.  These runs will be discussed in section \ref{sec:resolution}.  See Table~\ref{tab:sims} for the parameters of all runs.

\begin{deluxetable}{llcccc}
\tablecolumns{5}
\tablewidth{0pt}
\tablecaption{Simulation Parameters
\label{tab:sims}}
\tablehead{
\colhead{Simulations}&\colhead{$N_{\rm root}$} & \colhead{$\alpha$}&
\colhead{$f_{\rm cond}$} & \colhead{Figures}}
\startdata
1 & 64 & -1.2 & 0 & \ref{fig_resolution}, \ref{fig_classic}\\
2 & 128 & -1.2 & 0 & \ref{fig_Spitzer}, \ref{fig_resolution}\\
3 (standard) & 256 & -1.2 & 0 & \ref{fig_rho} - \ref{fig_M_T}\\
%\ref{fig_rho}, \ref{fig_dEdt}, \ref{fig_vr}, \ref{fig_time}, \ref{fig_compress}, \ref{fig_rotation}, \ref{fig_M_T}
4 & 128 & -0.2 & 0 & \ref{fig_resolution}\\
5 & 128 & -0.6 & 0 & \ref{fig_resolution}\\
6 & 128 & -1.2 & 0.1 & \ref{fig_Spitzer}\\
7 & 128 & -1.2 & 0.3 & \ref{fig_Spitzer}\enddata
%\tablecomments{ }
\end{deluxetable}

We also performed several runs with isotropic thermal conduction suppressed from the standard Spitzer value by different factors of $f_{\rm cond}$ in order to test its influence on the evolution of the cooling catastrophe (see Section \ref{sec:thermal} for details). Two different suppression values were tested ($f_{\rm cond}=0.1$ and $f_{\rm cond}=0.3$). We did not include star formation or any feedback mechanism in our simulations.  

In order to facilitate analysis of the simulation, Enzo generates an output when a new refinement level is reached for the first time.  This produces outputs throughout the collapsing phase up until the point when the maximum refinement level $l_{max}$ is achieved. To better resolve the evolution after this, starting from a few Myr before $l_{max}$ is reached, the simulation writes one output every $1/3$ Myr and continues to do so afterwards. We run the simulation until the estimated energy from AGN feedback is more than strong enough to offset cooling, which occurs approximately 10-20 Myr after reaching $l_{\rm max}$.

For completeness, we note that these simulations were run using comoving coordinates, but over a sufficiently brief span ($\Delta z < 0.2$) that expansion did not play a role.   We adopted cosmological parameters corresponding to $H_0 = 50$ km s$^{-1}$ Mpc$^{-1}$, $\Omega_\lambda = 0$ and $\Omega_m = 1$, but note that the cosmological parameters are irrelevant to the physics inside the self-gravitating cluster and thus have no significant influence on its evolution or on the initial condition (they just serve to set the internal code units). The simulation starts at $z = 1$, which is chosen to give enough time for the cluster to evolve.

% --------------------------------------------

\section{Results}\label{sec:results}

In this section we present and discuss the results from our standard simulation.  In Section \ref{sec:results_evolution}, we describe the time evolution of the cooling gas. In Section \ref{sec:results_catastrophe}, we describe when and how the cooling catastrophe happens.

% --------------------------------------------

\subsection{Cluster Evolution}\label{sec:results_evolution}

We show in Figure~\ref{fig_rho} the evolution of the gas density of our standard simulation with $N_{\rm root} = 256$. Throughout the cluster core ($r \lesssim 30$ kpc) the density slowly increases and the temperature decreases over the first few hundred million years.  The density in the center ($r \lesssim 100$ pc) grows more compared to the outer part of the core region $r \sim 0.3 - 10$ kpc, where the temperature shows a plateau. After about 300 Myr, which corresponds to the cooling time of the gas in the initial conditions, the evolution rapidly accelerates, and the density increases by several orders of magnitude within a million years.  Notice that the outputs are not evenly spaced in time (output times are marked and color-coded in the upper-right panel of Figure~\ref{fig_rho})  This marks the onset of the cooling catastrophe, and occurs in the central region of the cluster.  We refer to the outer boundary of the region where this catastrophe happens as the ``transition radius''. This is different from the traditional cooling radius which is defined as the radius at which $t_{cool}$ is equal to the Hubble time (and which occurs near 100 kpc). The huge jump of the central density from $\sim 10^{-23}$ g cm$^{-3}$ to $\sim 10^{-21}$ g cm$^{-3}$ happens within only a short period of time, about $3$ Myr, which also corresponds well with the drastic decrease in the central temperature seen in the lower-left panel of Figure~\ref{fig_rho}.  After the cooling catastrophe, the transition radius steadily grows in time, although points exterior to the transition radius evolve only very slowly.

\begin{figure}
\begin{center}
\includegraphics[scale=0.4]{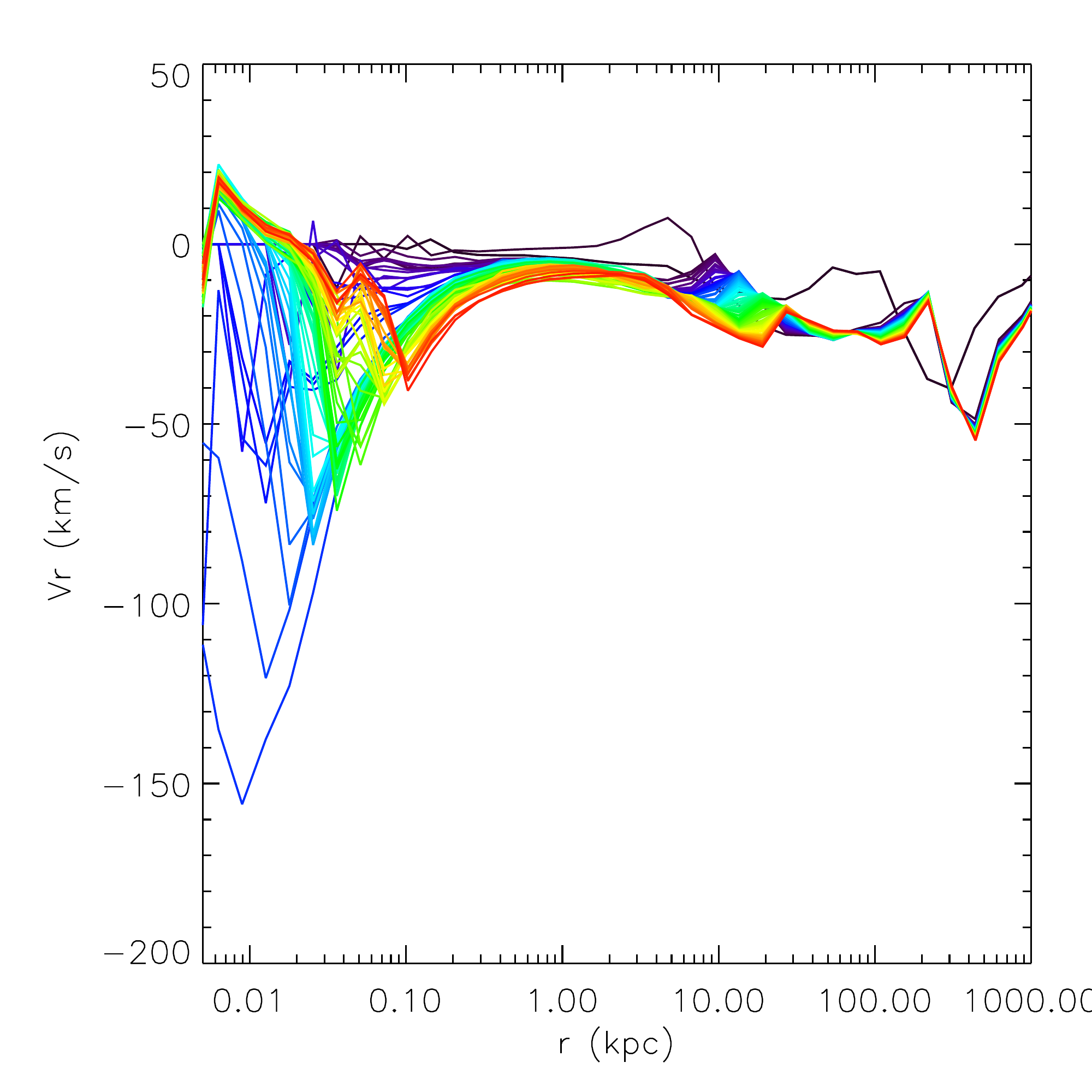}
\caption{Radial velocity of the gas (spherically averaged), color-coded for the same times as in Figure~\ref{fig_rho}. 
\label{fig_vr}}
\end{center}
\end{figure}

The pressure profile (Figure~\ref{fig_rho}, lower right panel) shows relatively little variation over time, at least in the outer parts. As more gas flows to the cluster center, the pressure slowly builds up, although the slope in the intermediate region (between 100 pc and 10 kpc) is quite shallow. As the cooling catastrophe occurs, the pressure in the very center rises quickly. A small region with a slightly positive pressure gradient forms later right at the inner edge of the transition radius (at $r \sim 50$ pc), creating a slight pressure ``hole" (or inversion). This pressure hole does act to drive some gas inflow, however, the pressure profile is remarkably constant at the transition radius, while the density and temperature change by orders of magnitude.  Both the size and the depth of the pressure hole are sensitive to the resolution of the simulation, as we will discuss in Section \ref{sec:resolution}, becoming less prominent as the resolution increases.  After the cooling catastrophe, the pressure profile is nearly time-independent.

\begin{figure*}
\begin{center}
\includegraphics[scale=.9]{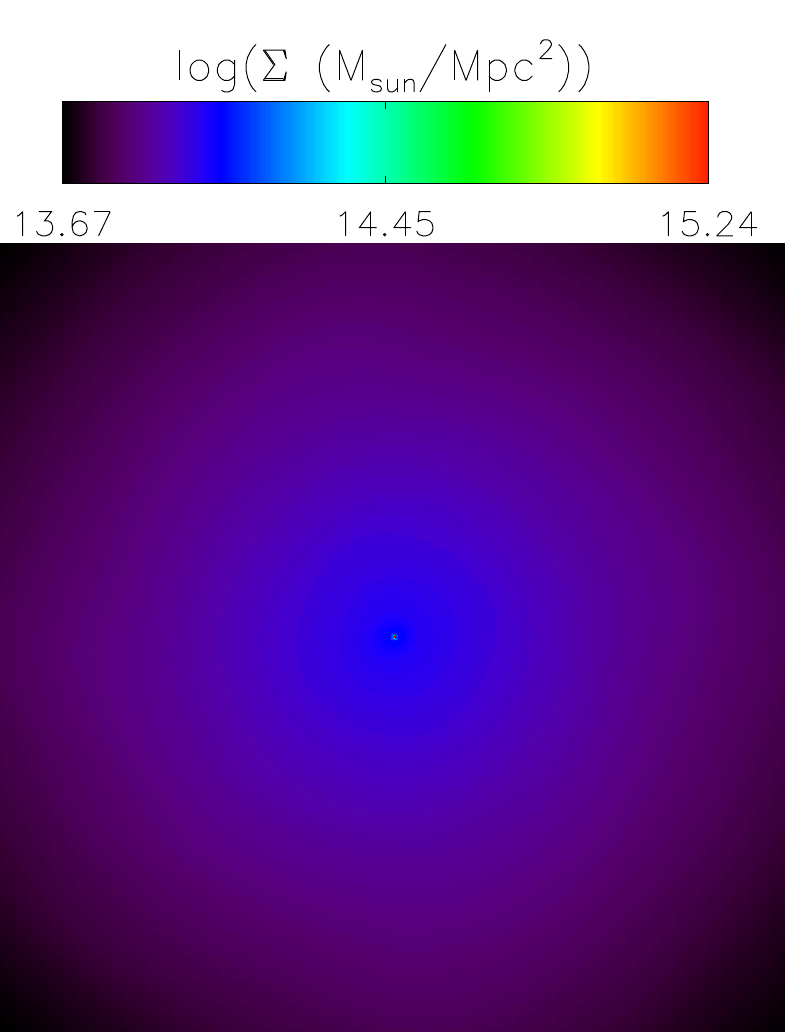}\includegraphics[scale=.9]{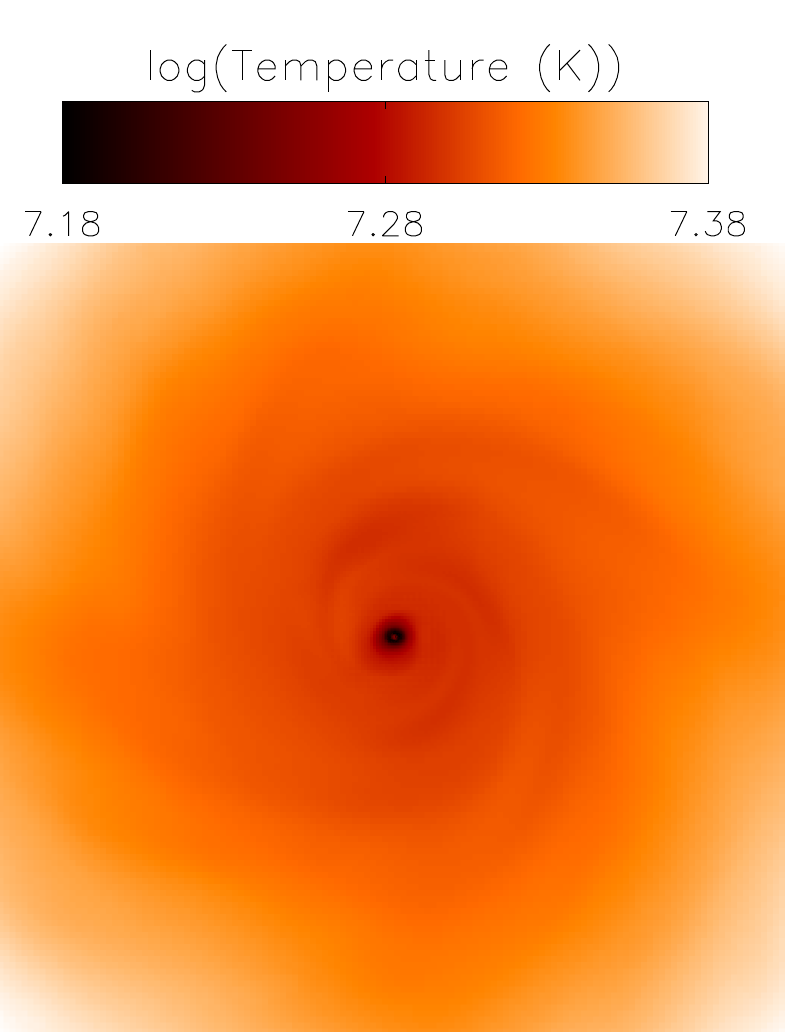}\\
\includegraphics[scale=.9]{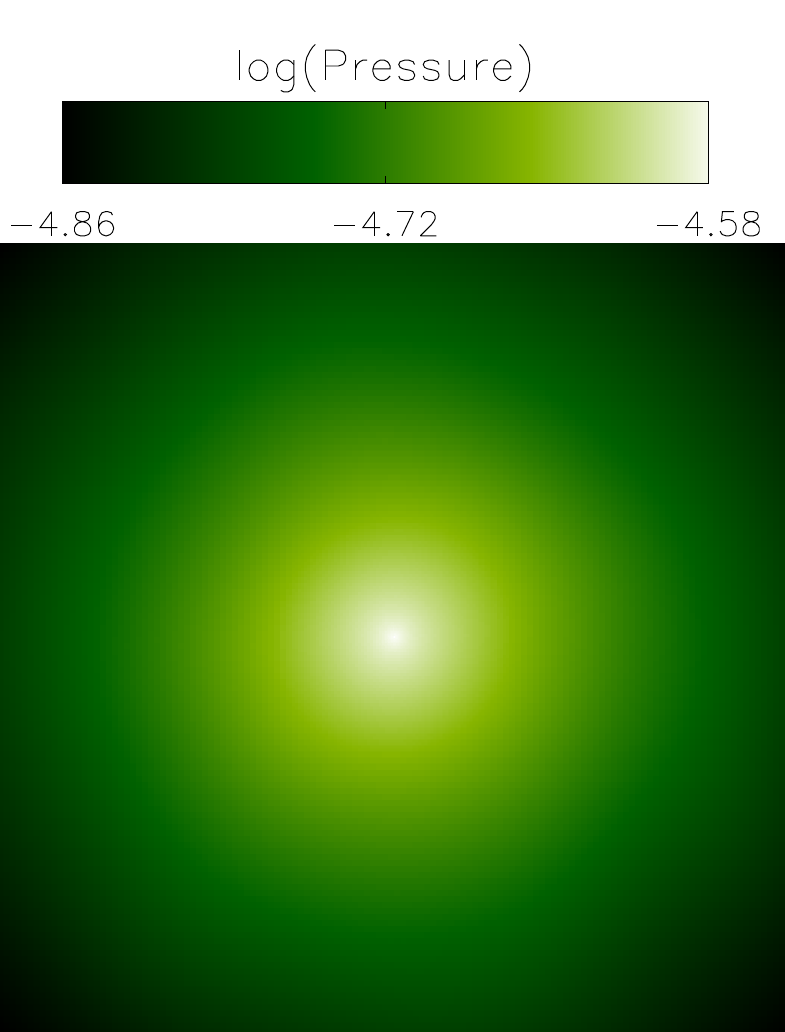}\includegraphics[scale=.9]{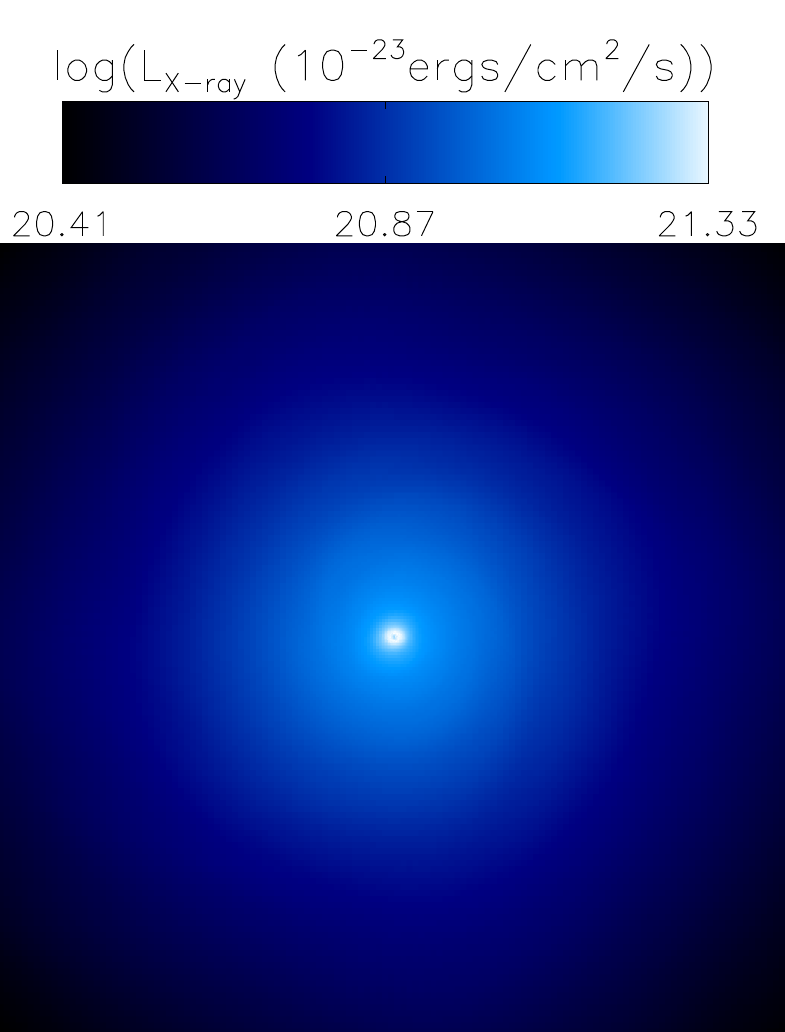}
\caption{The gas surface density (upper left), X-ray weighted temperature (upper right), pressure (lower left) and the X-ray luminosity ($0.5 - 2\;$keV) (lower right) in the cluster core projected along the direction of the initial angular momentum z-axis at late stage of the simulation, about 296 Myr after the initial time. The size of the region is roughly ($16.6$ kpc)$^3$.
\label{fig_project1}}
\end{center}
\end{figure*}

It is clear from Figure~\ref{fig_rho} that the cluster can be divided into three regimes in radius. In the outer region ($r>20$ kpc), where the NFW dark matter dominates the gravitational potential, the gas properties stay relatively constant during the simulation.  Inside the core, but outside the transition radius ($0.1$ kpc $\lesssim r\lesssim 20$ kpc), where the gravitational potential is dominated by the BCG, the density and pressure increase slowly while the temperature profile is nearly flat.  Finally, in the very center of the cluster ($r \lesssim 0.1$ kpc), within the radius of influence of the SMBH where the gravity is dominated by the SMBH ($\sim 50$ pc, see Figure~\ref{fig_g}), the cooling catastrophe first occurs at small radius. Then the transition radius moves slowly outwards with time, from $\sim 10$ pc when the collapse first happens to about $\sim 100$ pc, about $10$ Myr later. 

Figure~\ref{fig_vr} shows the evolution of the radial inflow velocity of the gas. Initially it is zero, and in the absence of cooling stays that way except in the outer region ($r > 100$ kpc), where the cluster is not in perfect hydrostatic equilibrium\footnote{This occurs because we use observationally-motivated density and temperature profiles which are not in perfect hydrostatic equilibrium.  See Section~\ref{sec:nonCC} for a discussion.}.  As radiative cooling proceeds, a slow but steady inflow develops.  This is slow ($\sim 20$ km/s) and nearly constant over a large range in radius, from 20 kpc to a few hundred pc.  Notice that this is not a steady state cooling flow because a constant velocity implies an accretion rate that rises with radius -- we discuss this in more detail below.  As the cooling catastrophe takes hold (recall that the outputs are not evenly spaced in time), the inflow velocity in the central region rises up to a few hundred km/s inside 100 pc, and then drops after an accretion disk forms.

So far, we have focused on one-dimensional analysis of the cluster evolution.  In fact, the evolution is remarkably symmetrical, despite the fact that we seed the initial conditions with substantial small-scale velocity motions which quickly generate density and temperature perturbations.  We show projections of the density, (X-ray weighted) temperature, pressure and X-ray luminosity of the central 16.6 kpc of the cluster in Figure~\ref{fig_project1}.  These are shown 296 Myr after the initial time, shortly ($\sim 16$ Myr) after the cooling catastrophe occurred.  Clearly the initial seeds have damped and the profiles are remarkably uniform.  A close examination of the temperature image reveals faint spiral structures due to the slight rotation imprinted in the initial conditions.  This demonstrates that the flow does not exhibit local thermal instabilities, as predicted by linear perturbation theory \citep{Malagoli87, BS89}.  

The primary instability (or catastrophe) which grows is a global one, as discussed above.  This can be seen as the increase in density in the very center of the projected image (and the decrease in the temperature).  The pressure map is quite smooth.  The X-ray map in Figure~\ref{fig_project1} shows a small rise in the central emissivity in this region, but it can be appreciated that this would be difficult to observe.  We return to the observability of this lower temperature gas in Section~\ref{sec:obs}.

We stop the simulation about 20 Myr after the cooling catastrophe first occurs because the estimated energy output from AGN feedback rapidly exceeds the energy loss through radiation.  We estimate that the feedback energy from the AGN as
\begin{equation}
\dot{E}_{\rm AGN}=\epsilon \dot{M}_{400} c^2  \;\;,
\label{eq-AGN}
\end{equation}
where $\dot{M}_{400}$ is the mass accretion rate measured at $r = 400$ pc, which is just outside the region dominated by the potential well of the SMBH, and $\epsilon$ is an efficiency parameter discussed in more detail below.  We stop the simulation shortly after this feedback rate exceeds $L_{\rm X50}$, which is the total calculated X-ray luminosity from the cooling gas inside $r<50$ kpc.

The parameter $\epsilon$ is the efficiency relating this mass accretion rate to the feedback energy that heats the ICM, and is really the product of three somewhat poorly constrained efficiencies.  The first is the fraction of $\dot{M}_{400}$ which actually accretes on to the black hole.  Because we measure this value quite close to the SMBH-dominated region (at 400 pc), we argue that a significant fraction (at least 10\%) of this mass will accrete on to the black-hole.  The second unknown efficiency is that associated with the conversion of accreted mass into energy, which we take to be approximately 10\%.  Finally, because the accretion rate is likely to be sub-Eddington, this feedback will be so-called `radio-mode', and we assume that a large fraction of the feedback energy ends up heating the ICM.  For the product of these three efficiencies, we take a conservative value of $\epsilon = 0.01$, although we recognize that the final value is uncertain.

\begin{figure}
\begin{center}
\includegraphics[width=0.5\textwidth]{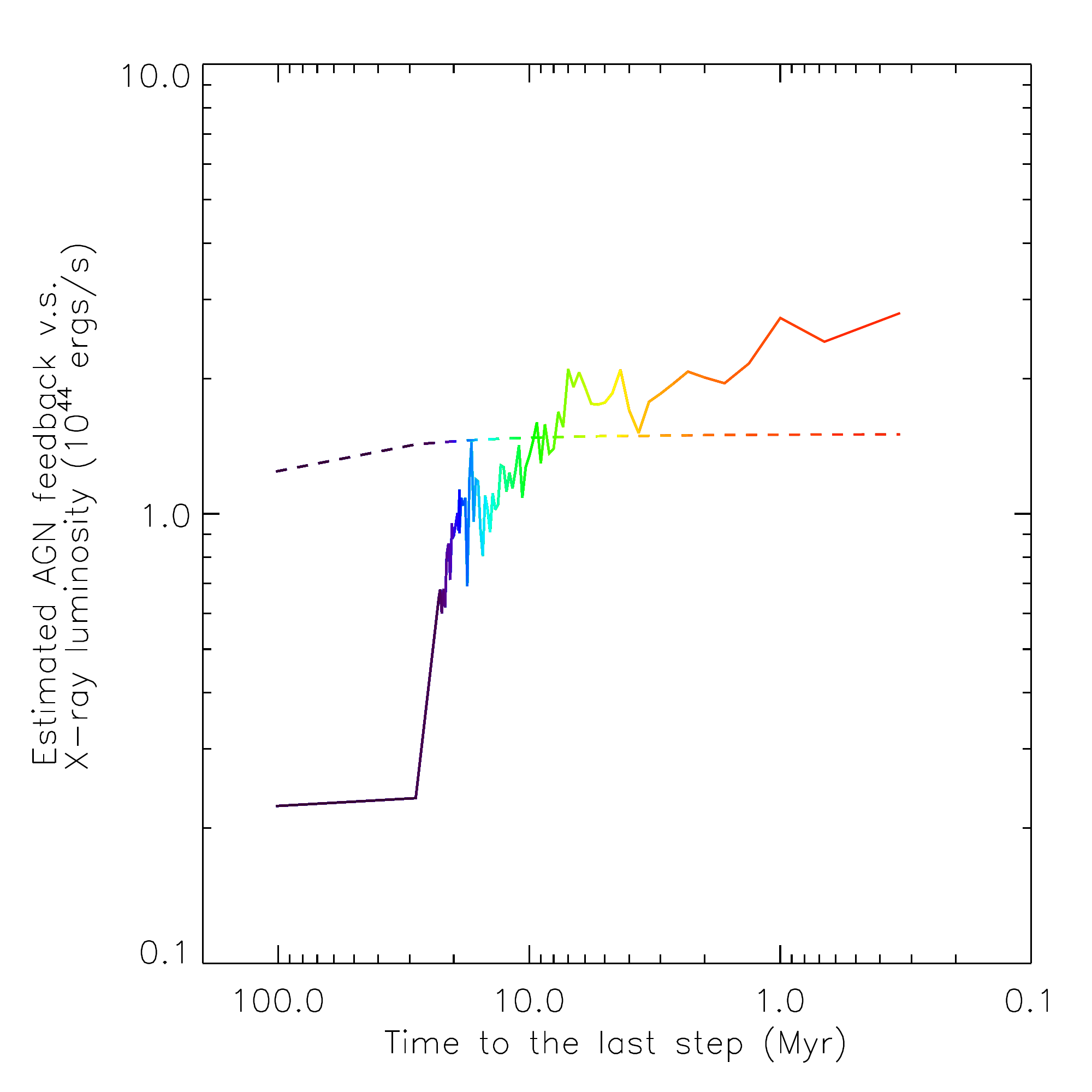}
\caption{The estimated ICM heating rate from the AGN (computed with Equation~\ref{eq-AGN}, shown as the solid line), compared with the total X-ray luminosity in the central $50$ kpc region (dashed line).
\label{fig_dEdt}}
\end{center}
\end{figure}

As shown in Figure~\ref{fig_dEdt}, as the gas inflow increases, $\dot{E}_{AGN}$ increases and exceeds $L_{\rm X50}$ after the cooling catastrophe. At the last step, $\dot{E}_{AGN}\sim 2.6 \times10^{44} \;\rm{ergs/s}$ and $L_{\rm X50} \approx 1.5\times10^{44}\;\rm{ergs/s}$, which has increased only about 50\% compared to the initial $L_{X50}$. We stop our simulations at that point because AGN feedback, which is not included, would be important and we only want to focus on the onset of the cooling flow. The total X-ray luminosity of the cluster is $6.8 \times 10^{44}\;\rm{ergs/s}$ at the end of our simulation.

% --------------------------------------------

\subsection{The Global Cooling Catastrophe}\label{sec:results_catastrophe}

\begin{figure*}
\begin{center}
\includegraphics[scale=.4]{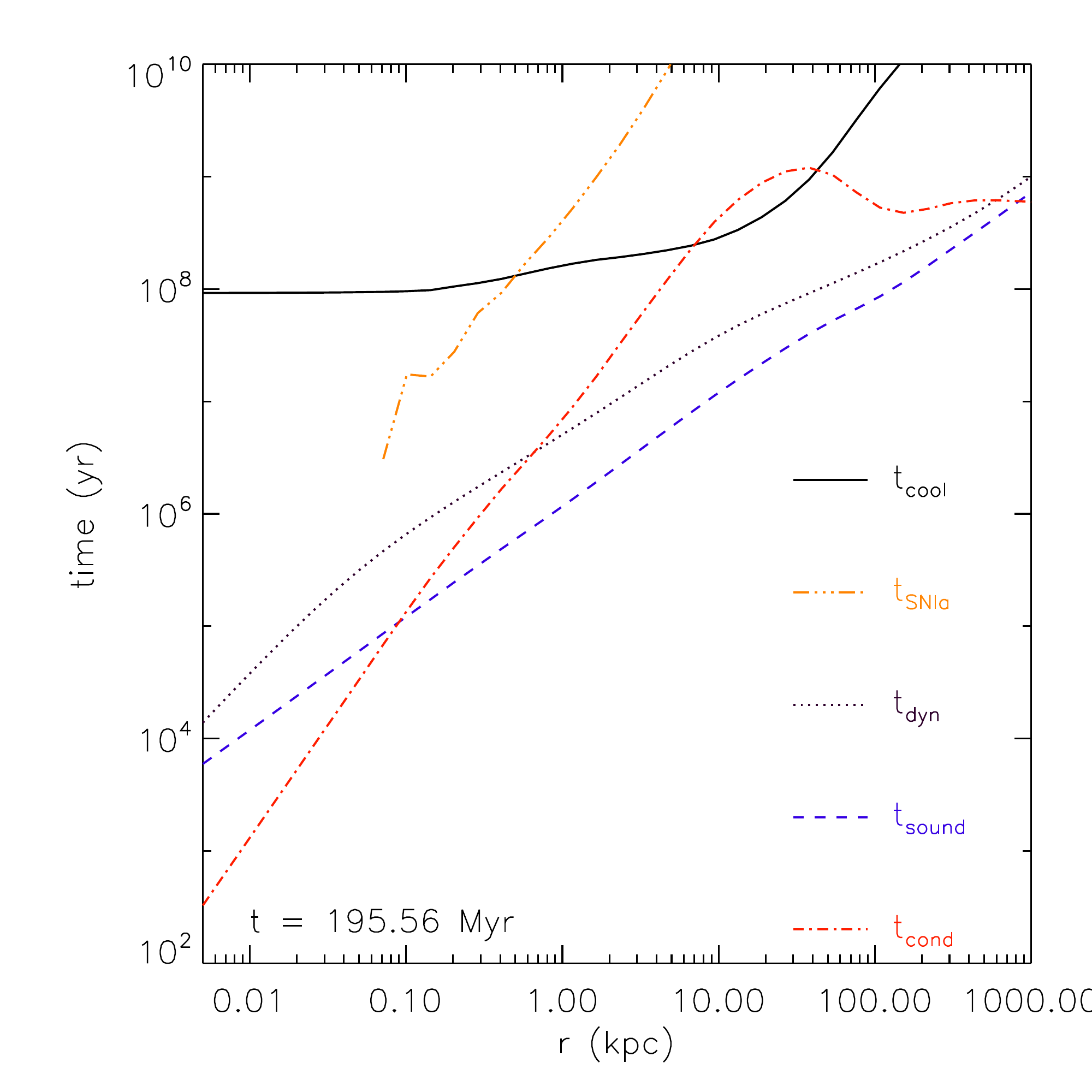}\includegraphics[scale=.4]{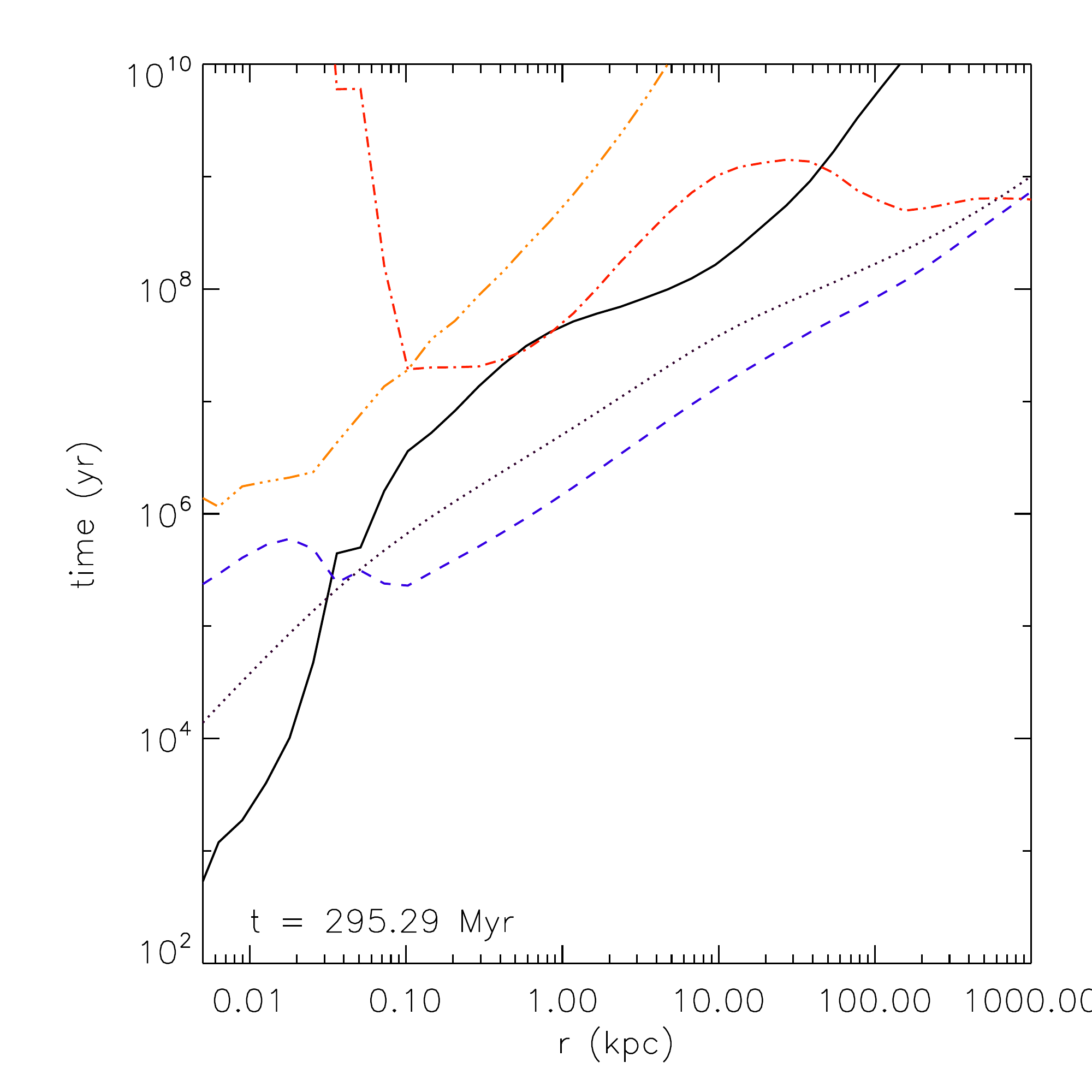}
\caption{Comparison of different time scales at different stage of evolution; see text for precise definitions. The thermal conduction time is calculated assuming a suppression factor of $f_{\rm cond}=0.3$. 
\label{fig_time}}
\end{center}
\end{figure*}

To better understand the physics behind the cluster evolution, we examine several relevant timescales in Figure~\ref{fig_time}. The dynamical time of a gaseous system can be defined as:
\begin{equation}
t_{\rm dyn} (r) = \frac{\pi}{2} \frac{r^{3/2}}{\sqrt{G M_{dyn}(r)}},
\end{equation}
where $M_{dyn}(r)$ is the enclosed mass of the system.

The system undergoes gravitational collapse when $t_{\rm dyn}$ is shorter than the sound crossing time
\begin{equation}
t_{\rm sound} (r) = \frac{r}{c_s},
\end{equation}
where $r$ is the radius of the region.

Figure~\ref{fig_time} shows the cooling, dynamical and sound crossing time scales before (left panel) and after (right panel) the cooling catastrophe happens. Initially the cluster is stable with $t_{\rm cool} > t_{\rm dyn} > t_{\rm sound}$. As the gas condenses and cools nearly isobarically in the center, the cooling time $t_{\rm cool}$ decreases while $t_{\rm sound}$ slowly increases (both at fixed radius). When $t_{\rm cool}$ drops below $t_{\rm dyn}$, which first occurs around $r \lesssim 50$ pc, the gas undergoes a full-fledged cooling catastrophe. At the same time, $t_{\rm sound}$ exceeds $t_{\rm dyn}$, making the core region gravitationally unstable and causing the gas to collapse to the cluster center under gravity.  Meanwhile, the gas can no longer maintain pressure equilibrium since $t_{\rm cool} < t_{\rm sound}$.  Also shown in Figure~\ref{fig_time} are the thermal conduction and the Supernovae Ia (SNIa) heating timescales which will be discussed in Section \ref{sec:thermal} and Section \ref{sec:Ia}, respectively.

This analysis leaves open two questions.  First, why does the gas temperature remain nearly constant in the intermediate region, between the transition radius (at a hundred pc) and the start of the cool core (at $r \sim 20$ kpc)?  Second, what supports the gas inside the transition radius?

\begin{figure}
\begin{center}
\includegraphics[width=0.5\textwidth]{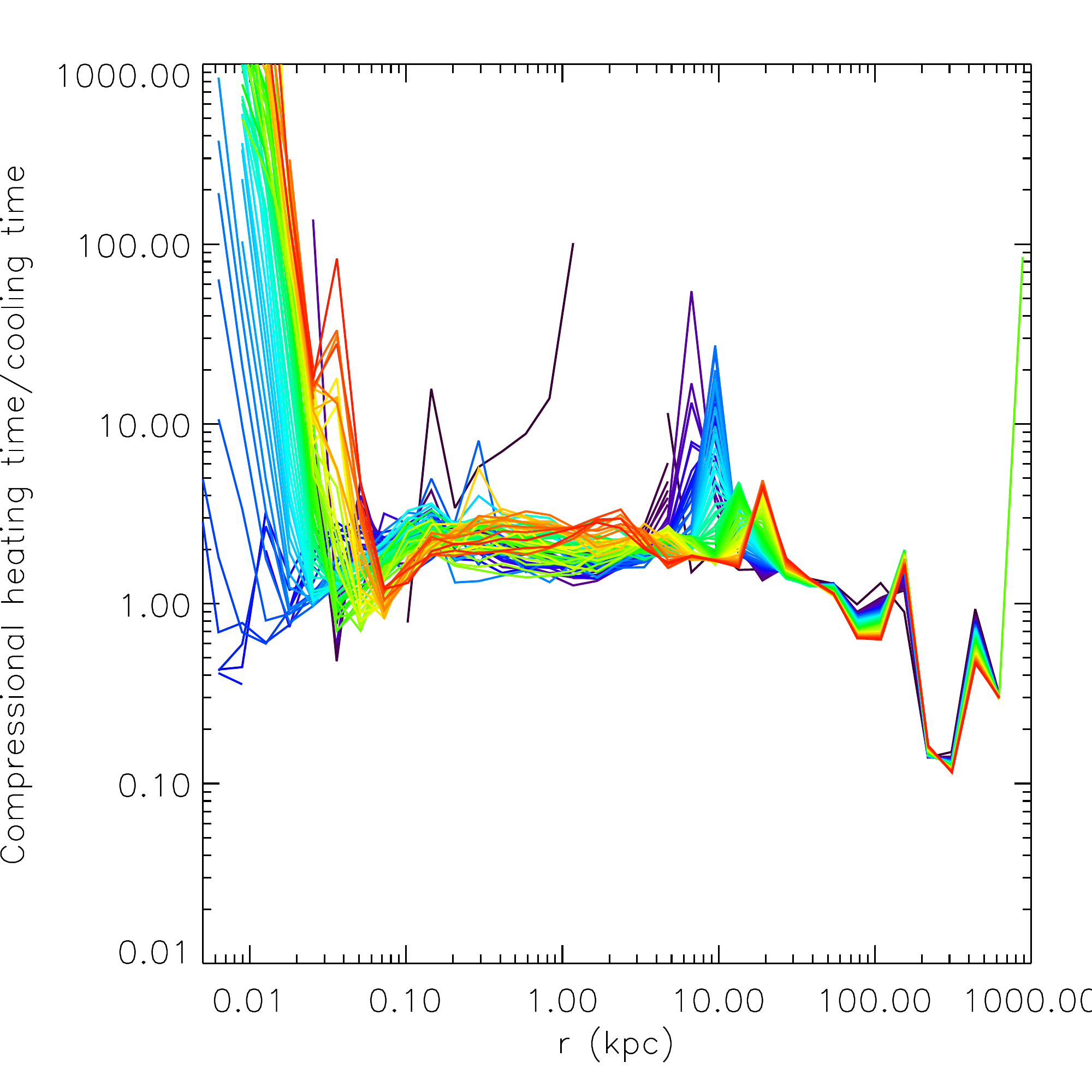}
\caption{The ratio of the (radial) compressional heating timescale over cooling time; see text for definitions. 
\label{fig_compress}}
\end{center}
\end{figure}

The first question arises because of the presence of the large temperature plateau noted earlier, which persists despite the fact that we evolve the system for many cooling times at these radii.  Clearly, the cooling must be balanced by some form of heating, and since the gas motions are subsonic, the only real candidate is compression heating.  We estimate the compressional heating timescale as: 
\begin{equation}
t_{\rm compress} (r) = \frac{1}{(\gamma - 1)\nabla \cdot v} \;\;,
\end{equation}
where $\gamma = 5/3$ and $\nabla \cdot v = \frac{1}{r^2} \frac{\partial (r^2 v_r)}{\partial r}$.  For simplicity, we only include the radial term, focusing on the role that the cooling flow plays in establishing the temperature plateau (we note that non-radial terms could decrease the heating time in the center, although manual inspection of the velocity indicates that the effect is minor).  We plot in Figure~\ref{fig_compress} the ratio of $t_{\rm compress}$ over $t_{\rm cool}$. At early times, over a large range of radius, $t_{\rm compress} \approx t_{\rm cool}$.  The inflow velocities which drive this heating are slight, a few tens of km/s over much of this range, as seen in Figure~\ref{fig_vr}, rising at small radius as the global cooling catastrophe sets in.

\begin{figure}
\begin{center}
\includegraphics[width=0.5\textwidth]{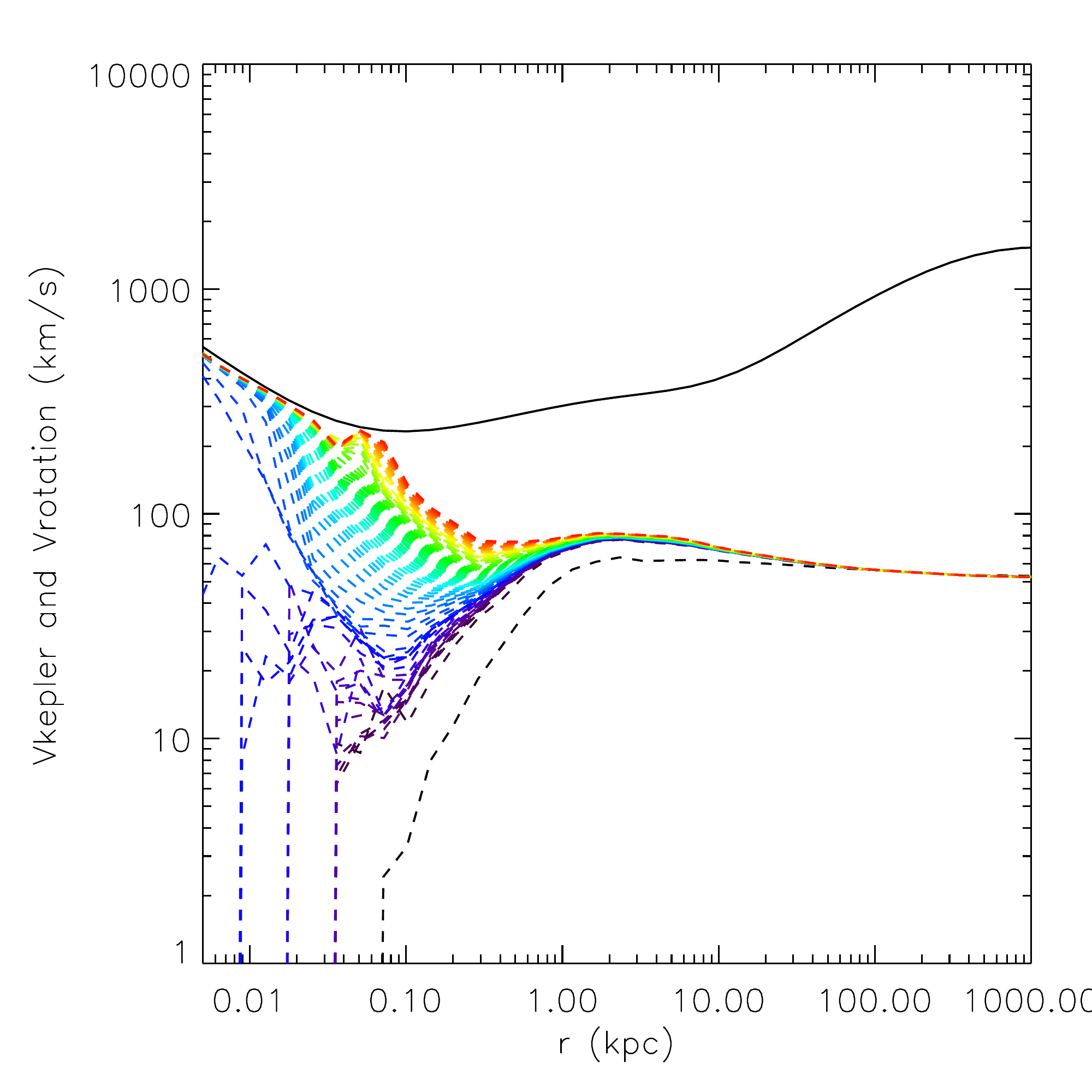}
\caption{The circular velocity of the gas (dashed lines) compared to the Keplerian velocity (solid line). The gas becomes rotationally supported in the very center at late times. 
\label{fig_rotation}}
\end{center}
\end{figure}

This brings us to the second question, the evolution of the gas within the transition radius.  As the cluster evolves, the cooling rate becomes higher in the center, requiring a larger inflow velocity to provide sufficient compressional heating. When this required velocity exceeds the sound speed, or if the gas becomes rotationally supported, compressional heating can on longer balance cooling and $t_{\rm compress}$ becomes larger than $t_{\rm cool}$ inside the transition radius.  If the inflow velocity grew to the sound speed, as would occur for a purely radial evolution, the transition radius would be identified as a sonic point, and inside the gas would freely fall toward the black hole.  We found this to occur in our lower resolution simulations (discussed in more detail below); however, in our best resolved, standard simulation, we find that the cold gas inside the transition radius forms a rotationally supported disk.  This is shown in Figure~\ref{fig_rotation}, which shows an estimate of the rotational velocity of the gas (computed by dividing the magnitude of the total specific angular momentum of the gas in a shell by the radius of that shell), compared to the Keplerian velocity.  At late times, inside the transition radius, the gas becomes rotationally supported.

\begin{figure*}
\begin{center}
\includegraphics[scale=.9]{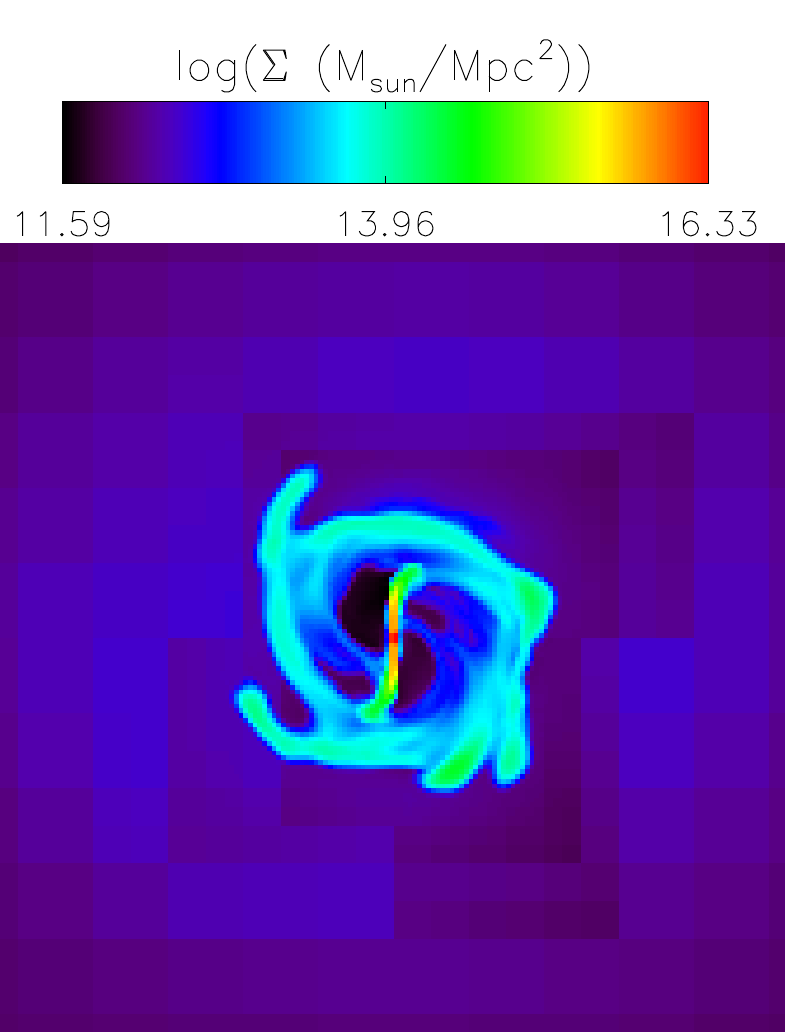}\includegraphics[scale=.9]{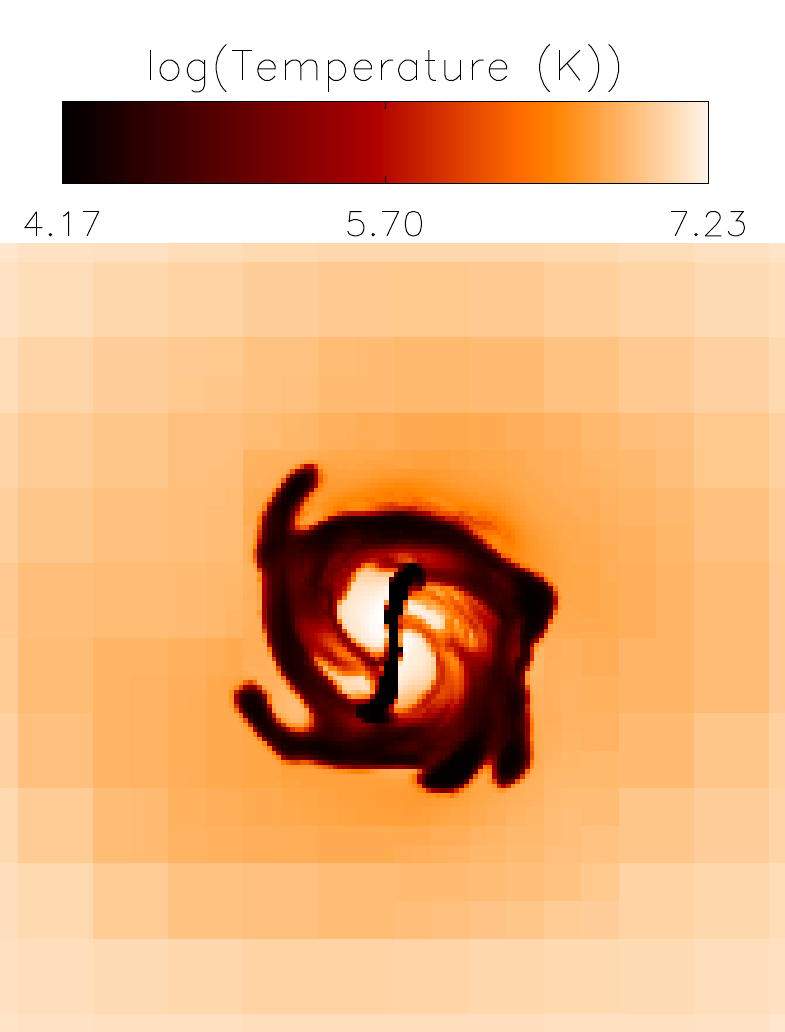}
\caption{The projected density (left) and density-weighted temperature (right) of a slice of gas in the very center of the cluster at the same time as in Figure~\ref{fig_project1}. The projection is along z-axis and the size of the region is about $330$ pc $\times 330$ pc $\times 16.6$ pc.
%same projected physical properties of the gas in the very center of the cluster: density (upper left), temperature (upper right), pressure (lower left) and the X-ray luminosity ($0.5 - 2\;$keV) (lower right). The time step is the same as in Figure~\ref{fig_project1}. The size of the region is about ($330$ pc)$^3$
\label{fig_project2}}
\end{center}
\end{figure*}

In Figure~\ref{fig_project2}, we show the projected density and density-weighted temperature in the central 330 pc for a slice of gas with a z-thickness of about $16.6$ pc.  This z-projection clearly shows the disk (x- and y-projections -- not shown here -- clearly demonstrate that this is a thin disk), which has a radius of about 50 pc.  In fact, in this particular run, we find an inner disk and an outer polar ring, which shows some sign of forming denser fragments; typical densities in the disk are of order $10^3$ cm$^{-3}$, but note that the disk is not well-resolved and so the detailed disk structure shown here should be treated with caution.

% --------------------------------------------------------------

\section{Discussion}\label{sec:discussion}

In this paper, we have presented results from the highest resolution simulation of the onset of cooling in a cool-core cluster.  We have demonstrated that the flow is remarkably uniform, with thermal instabilities not growing outside of the central few hundred pc, where the temperature drops rapidly at a point we have termed the transition radius.   In addition, we have shown that the flow natufrally generates a nearly constant temperature state outside of this transition radius.  Inside, a rotationally supported accretion disk forms around the central SMBH.   This time-dependent flow is not in steady state and is not, without additional heating, a solution to the cool core problem.  Nevertheless, we have made substantial progress in delineating exactly when and where cold, dense gas first condenses out of the flow.  

However, there are a number of unanswered questions, including a better understanding of why this solution occurs (section~\ref{sec:potential}), a detailed examination of the observational predictions of the the final simulation state (section~\ref{sec:obs}), a first attempt to examine the impact of thermal conduction (section~\ref{sec:thermal}) and Type Ia SN heating from stars in the BCG (section~\ref{sec:Ia}).  In section~\ref{sec:resolution}, we show that high numerical resolution is required to obtain these results, and with lower resolution, the transition radius first forms at much larger radius.  Finally, we argue that the results are robust to changes in the initial conditions (section~\ref{sec:nonCC}), and then compare these results to previous work (section~\ref{sec:comparison}).

% --------------------------------------------

\subsection{Cluster Profile}\label{sec:potential}

To better understand what determines the structure of the gas and why there are three regimes seen in the gas density, temperature and pressure profiles in Figure~\ref{fig_rho}, we carry out an approximate analytic analysis assuming hydrostatic equilibrium, which is valid before the cooling catastrophe happens (or at radii larger than the transition radius) when the inflow velocity $v_r \ll c_s$.

In hydrostatic equilibrium, the gas pressure $P(r)$ balances the gravitational acceleration $g(r)$
\begin{equation}
\frac{dP(r)}{dr} = - \rho(r) g(r) \;\;,
\end{equation}
where $P = {\rho k_b T}/{\mu m_H}$ and $\rho(r)$ is the gas density. This by itself is not sufficient to determine the density and temperature profiles uniquely; however, if we also use the fact, as suggested by Figure~\ref{fig_compress}, that compressional heating balances cooling, we can make more progress.  We write this as,
\begin{equation}
\rho g v_r \approx n^2 \Lambda(T)
\label{eq:balance}
\end{equation}
where $\Lambda(T)$ is the cooling rate and $v_r$ is the radial velocity.  We have assumed the work done on a fluid element as it flows inward is balanced entirely by radiative cooling.  Simplifying further, if we assume that $\Lambda(T) \approx \Lambda_0$ is independent of temperature, as is appropriate for gas around a few keV, in the cool region (see Figure~\ref{fig_rho}), and that the inflow velocity is nearly constant (see Figure~\ref{fig_vr}), then we can obtain an expression for the gas density: $n = \mu m_H g v / \Lambda_0$, and so $\rho \propto g$.   

We expect this result to hold primarily in the intermediate region of the flow, from the transition radius at about 100 pc to about 20 kpc.  This region is dominated by the potential of the BCG, which is reasonably well fit over this range with a power law $g(r) \sim  r^{0.75}$.  As predicted by the above argument, this is an excellent approximation to the density profile seen in Figure~\ref{fig_rho}.  If we combine this result with the equation of hydrostatic equilibrium, we predict that the temperature profile should be $T(r) \propto r^{0.25}$, which is also in reasonable agreement with the nearly flat temperature profile we find in Figure~\ref{fig_rho} over this radial range.

To see how changing the form of the gravitational potential in the cluster core can affect the structure of the cooling gas, we performed a test simulation with only the NFW dark matter but without the BCG and the SMBH and compare it with the standard run with the same number of cells on the root grid ($N_{root} =128$). The cooling catastrophe starts inside $r\sim 1$ kpc, much larger than the standard run. This is because the temperature decreases as $T \sim r$ towards the center throughout the core under the NFW gravitational potential and no temperature plateau forms.

% --------------------------------------------

\subsection{Observational Comparison}
\label{sec:obs}

One problem with the classic cooling flow model is that the expected amount of cool gas (i.e. gas with $T \lesssim {T_{\rm vir}}/3$, or $T \lesssim 2-3$ keV for massive clusters) is much higher than that observed in X-ray observations of cool core clusters. In our simulations, due to the small size of the region where the cooling catastrophe actually happens, the amount of cool gas is small and the X-ray luminosity does not change much in the core (see Figure~\ref{fig_dEdt}), which would explain the lack of cool gas observed in the X-ray. This can also be seen from the simulated X-ray luminosity (Figure~\ref{fig_project1}). Note that feedback is still required since this will not last without energy input: the amount of cool gas will grow with time and approaches the classic cooling flow prediction if we let the simulation run further (see Section \ref{sec:comparison}).

\begin{figure}
\begin{center}
\includegraphics[width=0.5\textwidth]{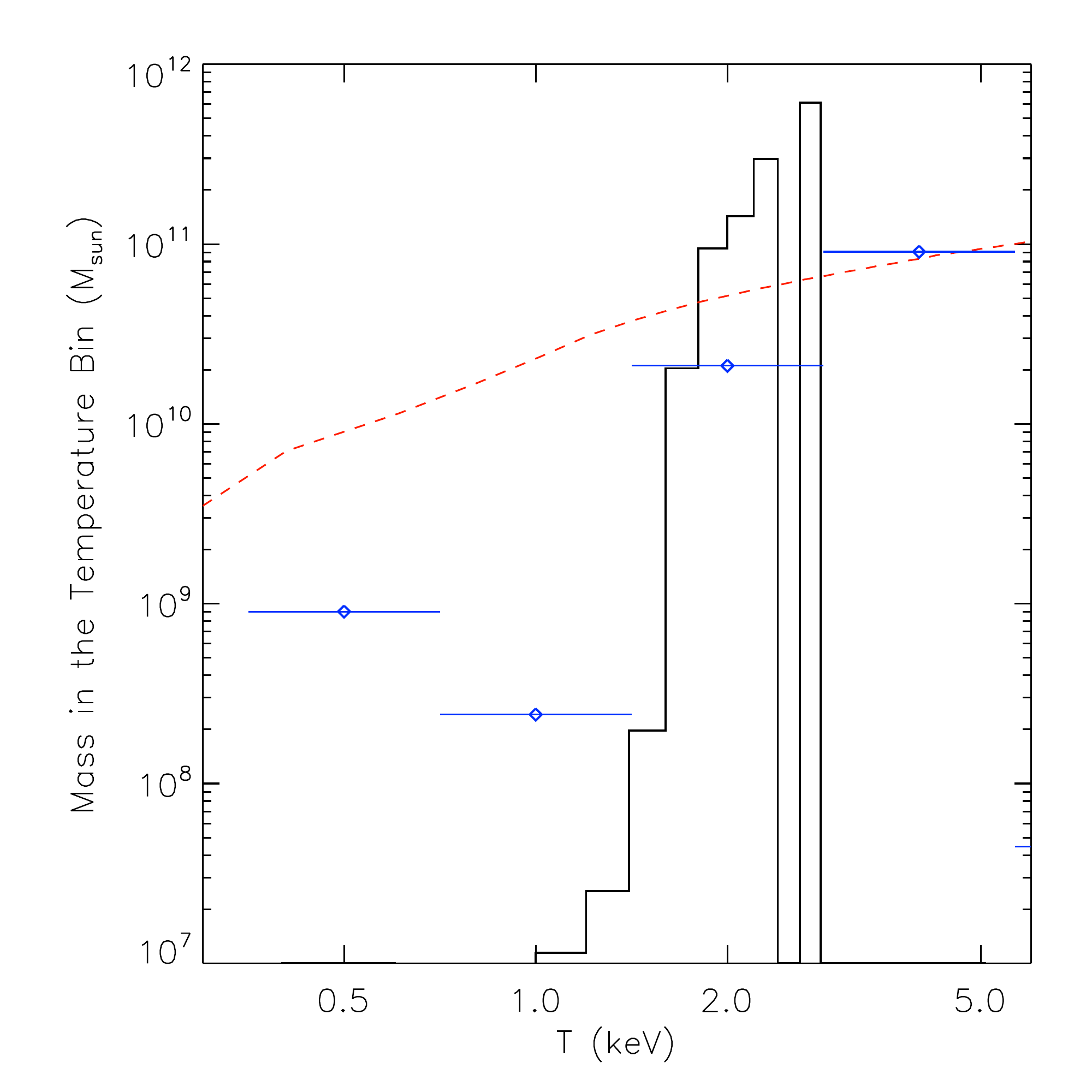}
\caption{Distribution of mass at fixed temperatures bins within $r < 32$ kpc. The black solid line shows the result from our simulation (296 Myr after the start, 16 Myr after the cooling catastrophe); the blue dots are the observed distribution within the innermost $1.5$ arcmin radius ($\sim 32$ kpc) and the red dashed line is the the classic cooling flow model prediction, taken from \citet{Fabian06}. 
\label{fig_M_T}}
\end{center}
\end{figure}

Since the observed temperature profile of galaxy clusters is usually obtained by fitting the X-ray spectra assuming a single temperature for the gas, it is hard for us to directly compare our temperature profile to those observed (since along a given line of sight, there exists gas with a range of temperatures). However, \citet{Fabian06} fitted the X-ray spectra of a long Chandra observation of Perseus with a multi-temperature model, allowing them to compute the amount of mass in each temperature range.  We compare the mass distribution from our simulation with their results in Figure~\ref{fig_M_T}, as well as with the prediction from the classic cooling flow model.  We find that the steep drop-off in mass between $1-2$ keV in our simulation is in good agreement with the observations and is much lower than the classic cooling flow prediction. 

Note that the recovery at $\sim 0.5$ keV in the observed mass distribution is due to the observed filaments (this also probably contributes to the 1 keV bin, see \citet{Fabian06}), which may be caused by the interaction between the cooling gas and AGN feedback that is not included in our simulations. The heating from AGN can potentially result in an increased amount of gas at $\sim 0.5$ keV, and we will study the impact of AGN heating in future work.  Additionally, there is an excess at around $3$ keV in our simulations and a lack of gas at around $5$ keV compared to observations. This is likely due to the initial condition we choose in our simulations: we start from a cool core configuration that matches Perseus' current configuration.  We suspect that if we started from slightly different conditions -- say an initial temperature plateau that was marginally higher, the fit could be improved.  In any case, we are not trying to fit Perseus in detail, but simply to make the point that the simulation naturally produces a dearth of low-temperature ($T \lesssim 2$ keV) gas.

% --------------------------------------------

\subsection{Thermal Conduction}\label{sec:thermal}

Thermal conduction can potentially delay or even suppress cooling instabilities. However, the existence of a magnetic field can decrease the efficiency of thermal conduction below the classical Spitzer value $\kappa_{\rm{S}}$ \citep{Spitzer56}.  This suppression can be parameterized by a factor of $f_{\rm cond}$ ranging from $0.1 \sim 1$ depending on the orientation and topology of the field \citep[e.g.][]{ZN03}. The characteristic timescale associated with thermal conduction can be estimated as 
\begin{equation}
t_{cond} (r) = \frac{\gamma}{\gamma-1}\frac{Pr^2}{f_{\rm cond}  \kappa_{\rm{S}} T} \;\;,
\end{equation}
where $\kappa_{\rm{S}} = 1.84 \times 10^{-5} T^{5/2}/\ln\lambda$ ergs s$^{-1}$cm$^{-1}$ K$^{-7/2}$ with the usual Coulomb logarithm $\ln\lambda \approx 37$.

Thermal conduction is more efficient along the magnetic fields than perpendicular to them. In the center of cool core galaxy clusters, where the temperature is increasing with radius, gas with even a sub-dominant magnetic field is susceptible to the heat-flux-driven buoyancy instability, or HBI \citep{HBI1, HBI2}.  This instability, in the saturated phase, orients the magnetic field to be perpendicular to the temperature gradient, and $f_{\rm cond}$ can be reduced to $\lesssim 0.1$ in the radial direction \citep{Parrish09}. On the other hand, the turbulent motion of gas can result in more entangled magnetic field, restoring conduction \citep{RO10}.  For simplicity, we assume isotropic conduction in our simulations, but we perform two additional simulations with the suppression factor $f_{\rm cond} = 0.1$ and $f_{\rm cond} = 0.3$, which can be thought of as corresponding to the HBI dominant case and the turbulent case. To save computational time, these simulations are done using $128^3$ root cells ($N_{root}=128$). 

We find that thermal conduction does not significantly affect the overall evolution of the cluster, except that it slightly changes the temperature and density profiles when the cooling catastrophe occurs, and it can somewhat change the timing of the cooling catastrophe itself by a few 10s of Myr.
% except that it may delay the collapse by a few Myr for $f_{\rm cond} = 0.1$ and larger for $f_{\rm cond} = 0.3$ -- we cannot be more precise because the stochastic influence of the initial random seeds have a similar timescale impact.  
This result is consistent with what one would expect from Figure~\ref{fig_time}: the thermal conduction timescale $t_{\rm cond}$ is shorter than the cooling timescale at $r \gtrsim 100$ kpc, where conduction can stabilize the local cooling instability, but at $r \lesssim$ a few tens of kpc where the cooling catastrophe first starts to develop, $t_{\rm cond}$ is longer than $t_{\rm cool}$. 

\begin{figure}
\begin{center}
\includegraphics[width=0.5\textwidth]{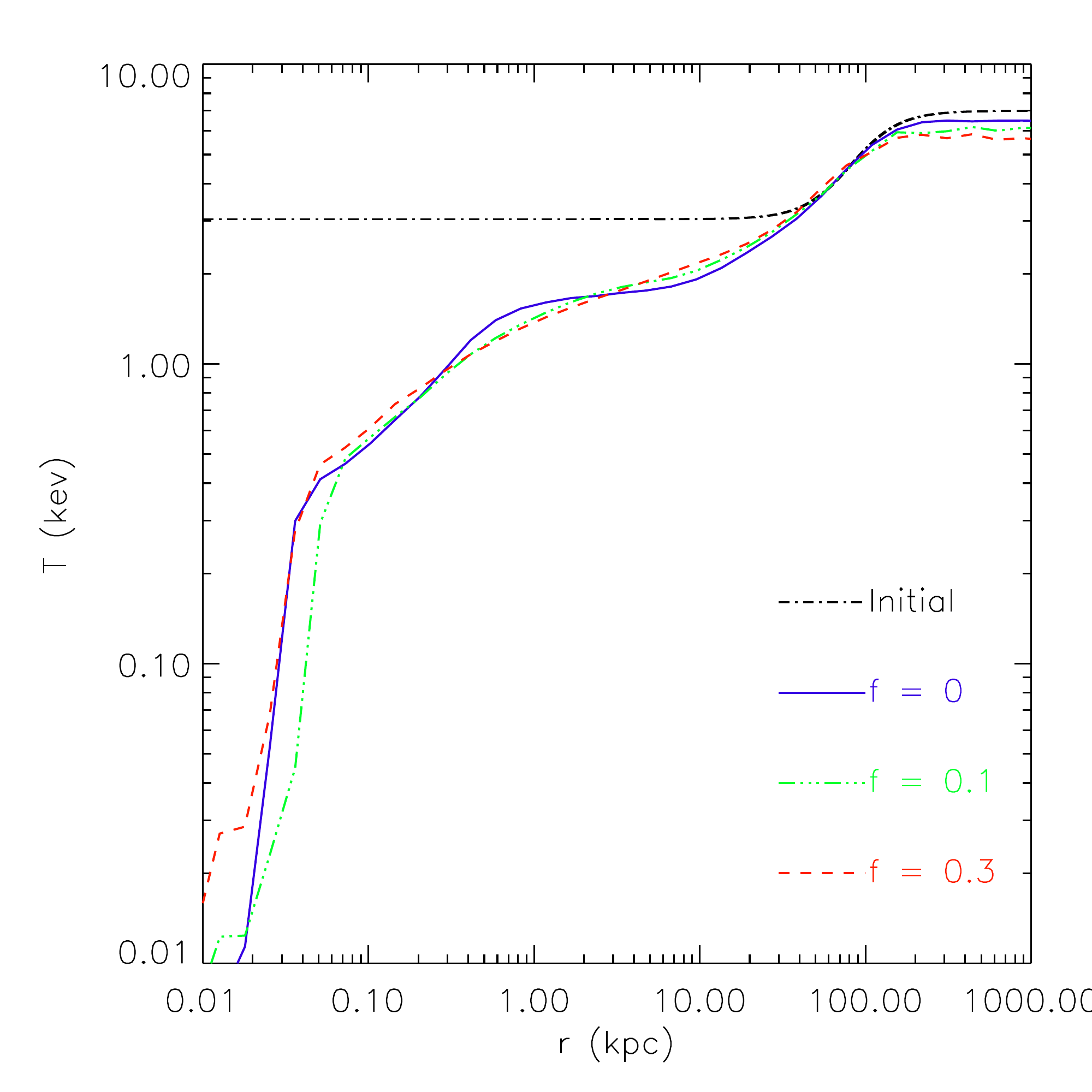}
\caption{Comparison of the temperature profiles of runs with different suppression factors $f_{\rm cond}$ of the heat conduction coefficient. Each profile is shown about 2 Myr after the cooling catastrophe first occurs (although the cooling catastrophe occurs at somewhat different times in each simulation).  The black dash-dotted line is the initial temperature profile (taken from observations of Perseus and extrapolated to the center). 
\label{fig_Spitzer}}
\end{center}
\end{figure}

To show the influence of thermal conduction near $r \sim 100$ kpc, we plot in Figure~\ref{fig_Spitzer} the temperature profiles from simulations with different suppression factors.  When $f_{\rm cond}$ is larger, the thermal conduction is more efficient, and thus the gas temperature is higher at $r \lesssim 100$ kpc but slightly lower at $r \gtrsim 100$ kpc because thermal energy is transported inwards from $r \gtrsim 100$ kpc where the temperature peaks. Inside the core region, the temperature evolution looks very similar with different $f_{\rm cond}$.  Note that the outputs in Figure~\ref{fig_Spitzer} are shown 2 Myr after the cooling catastrophe occurs in order to make a fair comparison in the central regions.  While the $f_{\rm cond} = 0$ and $f_{\rm cond} = 0.1$ runs collapse at almost exactly the same time ($t = 295$ Myr), the high conduction run actually collapses slightly earlier (about 20 Myr) because it damps the initial temperature perturbations caused by the relaxation at the beginning of the simulation which, in turn, are due to the slight deviation from hydrostatic equilibrium mentioned earlier.

% --------------------------------------------

\subsection{Supernovae Ia Heating}\label{sec:Ia}

The cooling catastrophe in these simulations takes place very close to the center of the BCG, where heating from Type Ia SN produced by stars in the BCG  may have an effect on the formation of cooling flows \citep[e.g.][]{Domainko04} and can be important especially in the central regions \citep[e.g.][]{Conroy}. We do not include supernovae heating in our simulations but here we try to analytically estimate their contribution. Type II supernovae can be ignored in the BCG due to its relatively old stellar population. The SNIa rate is $\sim0.1$ per $10^{10}\ \rm{M_{\odot}}$ per century in galaxy clusters \citep{Sharon07}. Assuming each SNIa releases $10^{51}$ ergs thermal energy into heating up the ICM, the heating rate then is expressed as M$_{stellar}$ (in $\rm{M_{\odot}}$) $\times 10^{38}$ ergs yr$^{-1}$ M$_{\odot}^{-1}$ and the SNIa heating timescale can be estimated as
\begin{equation}
t_{\rm SNIa} = \frac{\rho_{\rm gas}\times \frac{3}{2}\frac{k_b T}{\mu m_H}}{\rho_{\rm stellar} \times 10^{38}\ {\rm ergs\ yr}^{-1} \rm{M}_{\odot}^{-1}}  \;\;,
\end{equation}
where $\rho_{\rm gas}$ and $\rho_{\rm stellar}$ are the local gas and the stellar densities.

This timescale is plotted in Figure~\ref{fig_time}, which shows that although SNIa can be an important source of heating at $r\lesssim 1$ kpc and might slightly delay the formation of the cooling flow, it would not prevent the cooling catastrophe because $t_{\rm cool}$ is always shorter than $t_{\rm SNIa}$ at all radii.

% --------------------------------------------

\subsection{Resolution and Numerical Method}\label{sec:resolution}

\begin{figure*}
\begin{center}
\includegraphics[scale=.4]{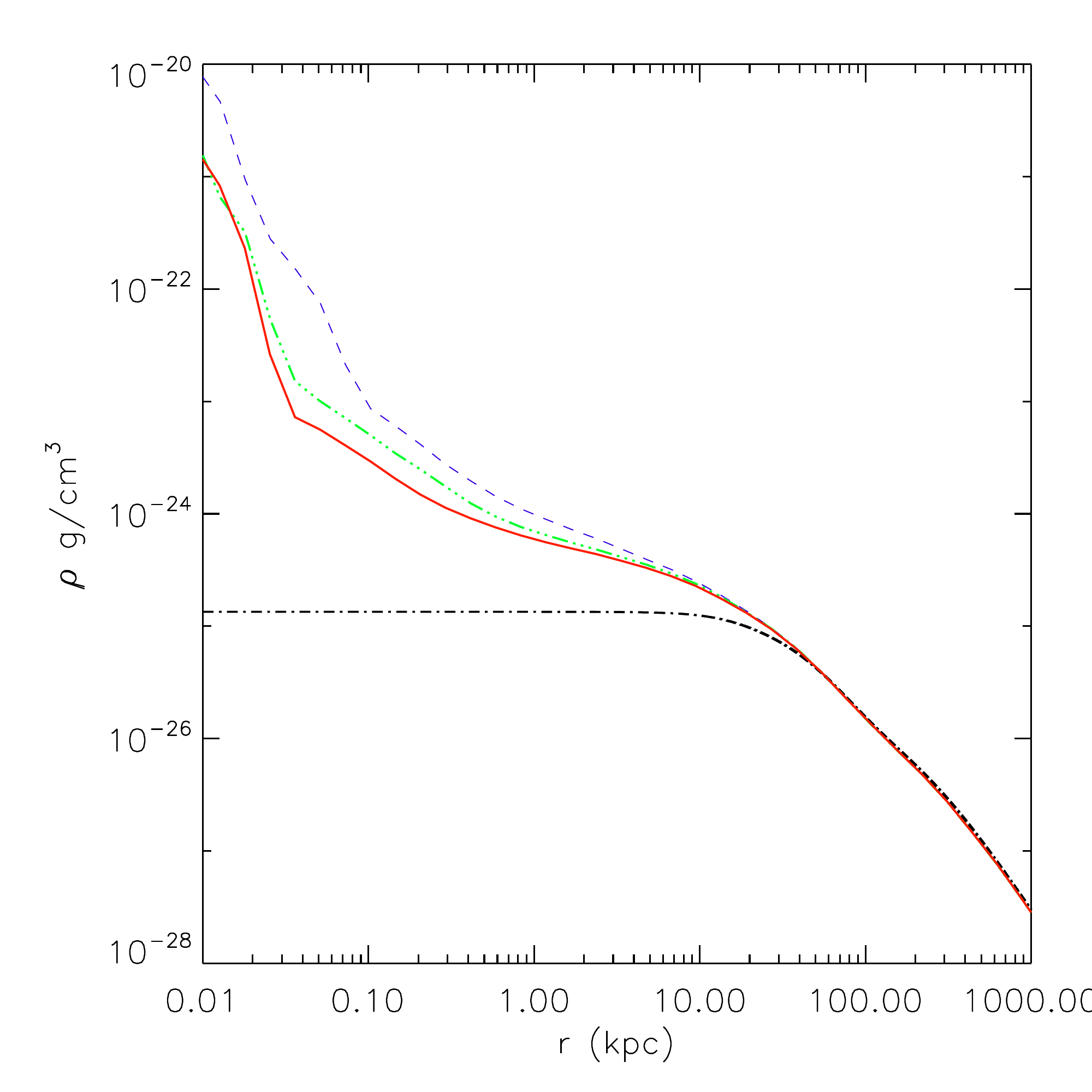}\includegraphics[scale=.4]{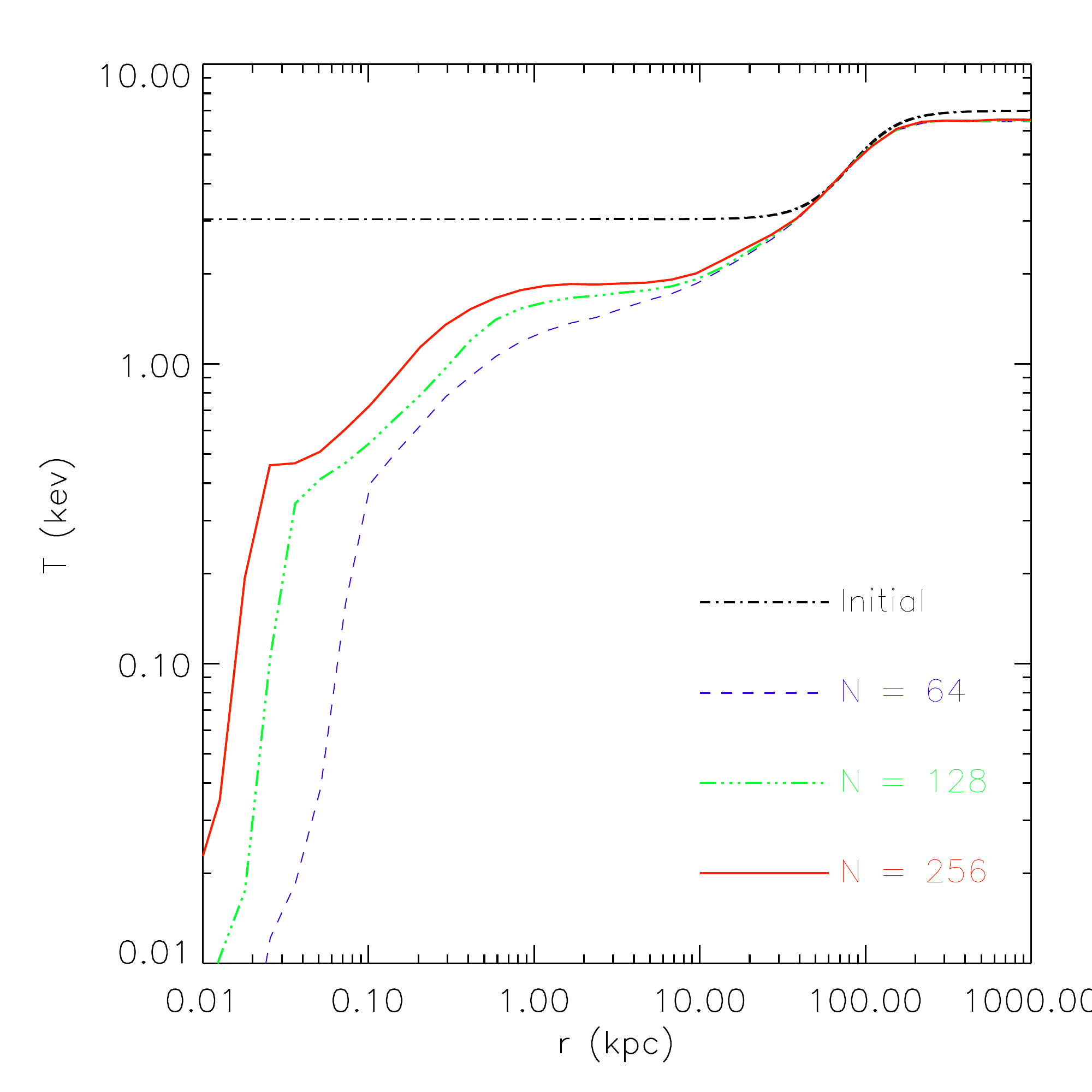}
\includegraphics[scale=.4]{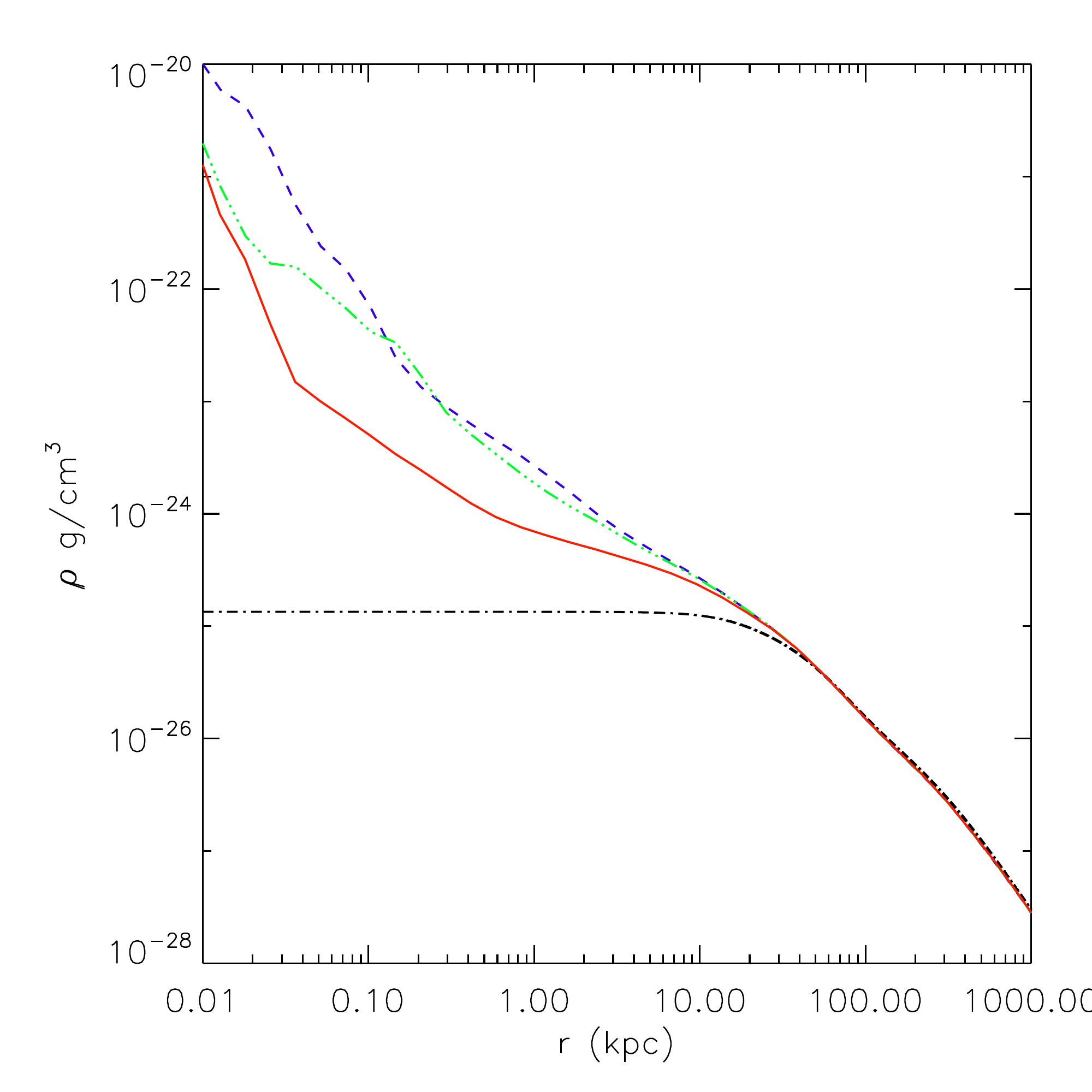}\includegraphics[scale=.4]{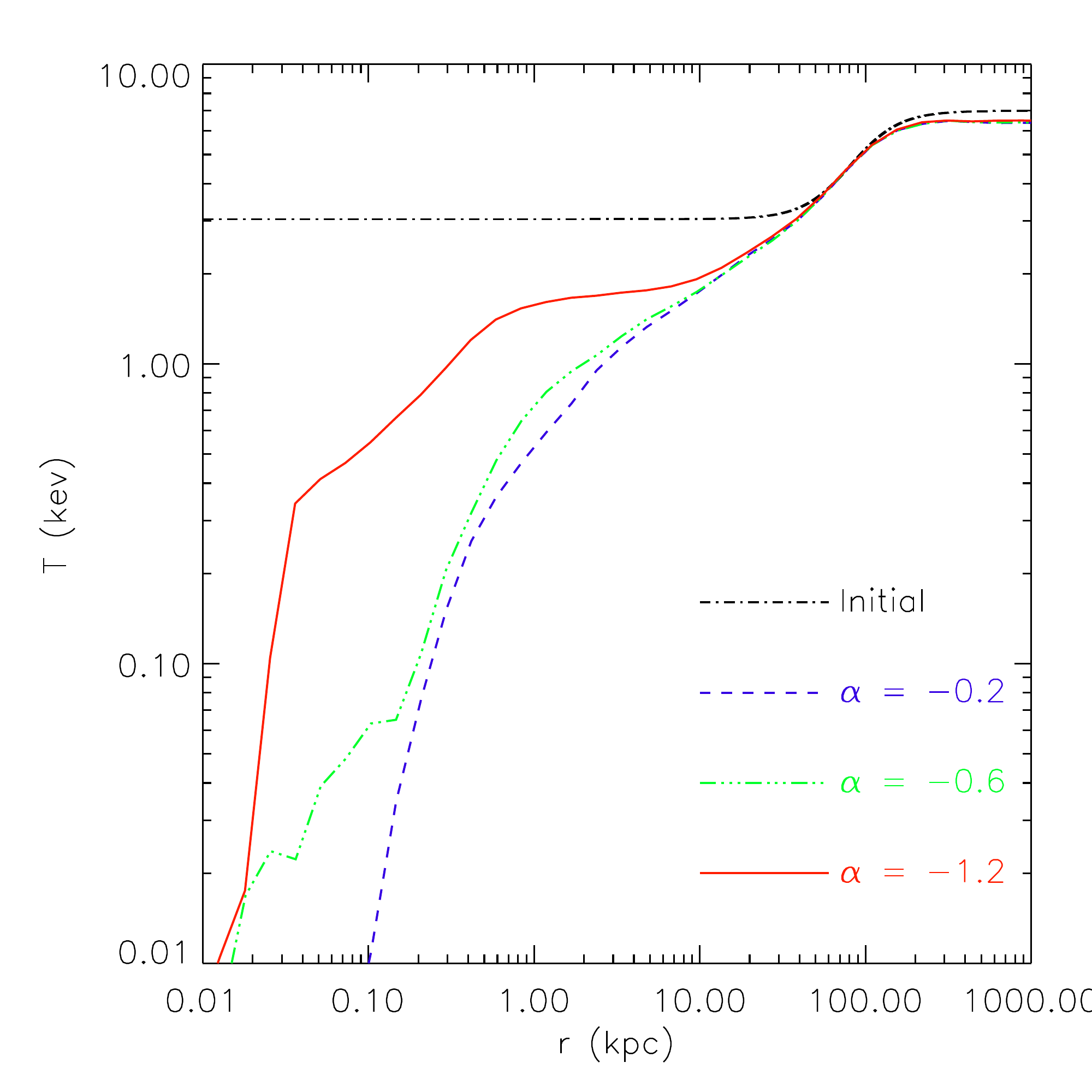}
\caption{The top two panels show the gas density (left) and temperature (right) approximately 2 Myr after the cooling catastrophe for three runs with varying root grid sizes, ranging from $N_{\rm root} = 64$ to $N_{\rm root} = 256$ (but each run with $\alpha = -1.2$). The bottom two panels show runs with $N_{\rm root} = 128$ but different $\alpha$ values (see Section~\ref{sec:resolution}), again for the same amount of time after the cooling catastrophe. 
\label{fig_resolution}}
\end{center}
\end{figure*}

To test the influence of resolution on our results, we perform simulations with $N_{\rm root} = 64$ and $128$ (keeping $\alpha = -1.2$) to compare with our standard $N_{\rm root} = 256$ run. Figure~\ref{fig_resolution} shows the density and temperature profiles of the cooling gas at a similar stage of the evolution ($\sim 2$ Myr after the cooling catastrophe starts).  The overall behavior of the solution is seen to be relatively independent of resolution; however, there is a systematic trend for the temperature plateau in the intermediate region (between 100 pc and 20 kpc) to be even flatter as the resolution increases, while the transition radius shrinks.  We do keep the maximum refinement level constant $(l_{\rm max} = 15$), so the highest resolution achieved increases by a factor of two in each case; however, even for the lowest resolution run, this corresponds to 7 pc, considerably smaller than the $\sim 100$ pc transition radius.  Therefore, we argue that it is not the maximum resolution that is the key, but the resolution achieved in the early stages of the collapse.

Since the point where the cooling catastrophe happens (the transition radius) is sensitive to the temperature and density profiles at earlier stages, we want to well resolve the early evolution of the cooling gas in the center of the cluster. Changing $\alpha$ in the baryon mass refinement criterion (see Section \ref{sec:methodology_refinement}) can affect the resolution, especially in the early stage inside the cluster core, and has a significant impact on the final results. The bottom two panels of Figure~\ref{fig_resolution} show a comparison of runs with the same number of root cells $N_{\rm root} = 128$, but different values for $\alpha$.  Again, the results are shown at a point roughly $2$ Myr after the cooling catastrophe starts. At fixed radius, the gas density is lower and the temperature is higher with more negative $\alpha$, and the transition radius is smaller. Noticeably, with $\alpha = -0.1$, the initial cooling region has a size of $\sim 1$ kpc, more than an order of magnitude larger than that in the run with $\alpha = -1.2$.  This is despite the fact that each calculation has the same maximum resolution.

We also notice that there is a degeneracy between $\alpha$ and $N_{\rm root}$: a higher $N_{\rm root}$ has a similar effect as a more negative $\alpha$. This is because they both result in better mass resolution (see Equation~\ref{eq:m_cell}) in the center of the cluster as the cooling catastrophe develops.  If we only increase the maximum refinement level $l_{max}$ without changing $N_{root}$ or $\alpha$, the result does not change despite a smaller cell size at the highest refinement level. Thus we argue that it is crucial to have sufficiently high resolution at the early stage of the gas evolution. 

Although we have not been able to achieve complete convergence in these calculations, it is clear that higher resolution tends to produce flatter temperature profiles and smaller transition radii.  

Another difference between simulations with different resolution is that in low resolution runs (e.g. with $N_{root} = 64$ and $\alpha = -0.2$), the pressure drops dramatically inside $r \lesssim 1$ kpc when the central gas density and temperature show a sudden change, forming a pressure hole which then grows deeper and larger with time. The pressure hole is deeper when the resolution is lower. The gas inflow velocity is larger than that in the high resolution runs and becomes supersonic at $r \sim 1$ kpc where the pressure gradient is the steepest, forming a sonic point and leading to a cooling catastrophe. Cold gas does not become rotationally supported as in the high resolutions runs but fragments inside the pressure hole via local cooling instability. 

Finally, we examine the impact of different methods for solving the hydrodynamics equations.  As noted earlier, we use the Zeus method for our simulations because of its fast performance, and robust treatment of cold regions. To test the accuracy of its results, we also carried out test runs using the piecewise parabolic method (PPM), which is third-order accurate with its high-order spacial interpolation. We used $N_{root} = 64$, $\alpha = -1.2$ and found results which were in good agreement with the Zeus solver.

% --------------------------------------------

\subsection{Impact of the Choice of Initial Conditions}\label{sec:nonCC}

In this section, we briefly examine how our results change as we vary details in the initial conditions.  We start our simulations assuming the gas is in hydrostatic equilibrium.  However, the NFW dark matter potential we use is fitted to the observed gas properties in the inner few hundred kpc region excluding the central $\sim 10$ kpc where the BCG starts to dominate the potential, and thus the resulting NFW parameters can vary slightly depending on the boundary radius we choose when fitting the parameters. Therefore the gas is not in perfect equilibrium when the simulation starts, especially in the outskirts and inside $\lesssim 10$kpc. Examination of the early evolution of the cluster shows that the temperature and density of the gas in the central kpc increases by about $30\%$ and $50\%$ respectively after about 20 Myr ($\sim t_{cross}$ at a few kpc).  They reach these values quickly, and then show little evolution until cooling becomes important on a few hundred Myr timescale.  In order to check if the initial transients are important for the cooling, we also carried out a simulation where we let the system settle down for $\sim 600$ Myr ($\sim t_{\rm dyn}$ at $r \sim 100$ kpc) without turning on cooling, so that the gas in the region of interest is in hydrostatic equilibrium when cooling starts.    The later evolution of the cluster is the same as the run that starts with the cooling on.  This indicates the transients are not playing a role in cooling; however, it does mean that the true "initial state" of the core region (i.e. the density and temperature profiles after the transients die out) is slightly different from our given initial profiles.  This is inevitable in the sense that the initial (observed) profiles are not in hydrostatic equilibrium.  To test the general robustness of our results, we have also experimented with slightly different sets of NFW parameters for the dark matter and find that the gas properties at the outskirts are slightly different with different NFW parameters, but the cooling flow evolution in the core region is not affected. 

In many galaxy clusters, there is an offset between the X-ray emission center and the BCG, although the offset tends to be smaller in CC clusters \citep[e.g.][]{Sanderson09}. To see if the offset would significantly change the results, we have performed one simulation with an initial offset of 20 kpc between the center of the gas and the gravitational potential. We find that the cluster gas settles down and re-centers on the BCG before the cooling catastrophe happens. This is consistent with the fact that $t_{\rm cool} > t_{\rm dyn}$ initially, and the cluster relaxes before cooling starts, and therefore the results do not differ from the simulations without the offset.  

To test the effect of other changes in our initial conditions, we performed one simulation without the initial random velocities. We find that changing the initial random velocity does not have a significant impact on the evolution of the cool core because the initial random velocity is damped before the cooling catastrophe happens. 

Since velocity perturbations do not directly perturb gas entropy, to further confirm that small scale perturbations do not grow outside the transition radius in our simulation, we performed a run with initial density perturbations instead of velocity perturbations. To do this, we multiplied the density in each cell in the initial conditions by a Gaussian factor with a mean of unity and a standard deviation of 10\%.  Again, we do not see the growth of any local instabilities.  This is in agreement with \citet{Joung11}, who found that perturbations in a hydrostatic atmosphere did not cool unless the perturbation was sufficiently non-linear that the cooling time in the perturbation dropped below the time for the clump to accelerate to the local sound speed (roughly the dynamical time).

We also performed a simulation without initial rotation. The gas in the very center in this run still eventually becomes rotationally supported because the random initial velocities eventually are amplified due to the conservation of angular momentum and a gas disk forms.  In fact, even in our standard run, a smaller disk along the x-axis forms inside the major disk along z-axis, which can be seen in Figure~\ref{fig_project2}. The size of the disk in the run without initial rotation is smaller at early times, and when the cooling catastrophe first occurs in that run, the gas in the very center has yet to become rotationally supported; however, the required inflow velocity to balance cooling has already exceeded the sound speed, and so the flow passes through a sonic point (in the run with initial rotation, rotational support occurs before a sonic point develops).  
%Note that even in the simulation we performed without initial rotation or random velocity, the disk still forms (with the initial size similar to the one with random velocity only) because random noise in the velocity field is inevitable. However, this does not make our result less robust since the random velocity in our standard runs mimic the turbulent motion in real clusters which overwhelms the numerical noise.

As noted earlier, the gravitational potential does play an important role in the cooling catastrophe and runs without a BCG produced significantly different results (see section~\ref{sec:results_catastrophe} for more details).

Finally, we carry out one simulation with a non CC configuration, where we use the same NFW dark matter profile for Perseus but set the initial temperature to be isothermal and compute the initial gas density assuming hydrostatic equilibrium. The initial $t_{\rm cool}$ in the center is about 2 Gyr.  Our simulation shows that after about 2 Gyr, the cooling starts to run away and a cooling flow develops in a way which is quite similar to what we see in the simulations with initial CC configurations. The temperature plateau is slightly different, which we think has to do with the difference in the initial gas to dark matter ratio.  We will return to this point in a later paper.

% --------------------------------------------

\subsection{Comparison with Previous Work}\label{sec:comparison}

\begin{figure*}
\begin{center}
\includegraphics[scale=.4]{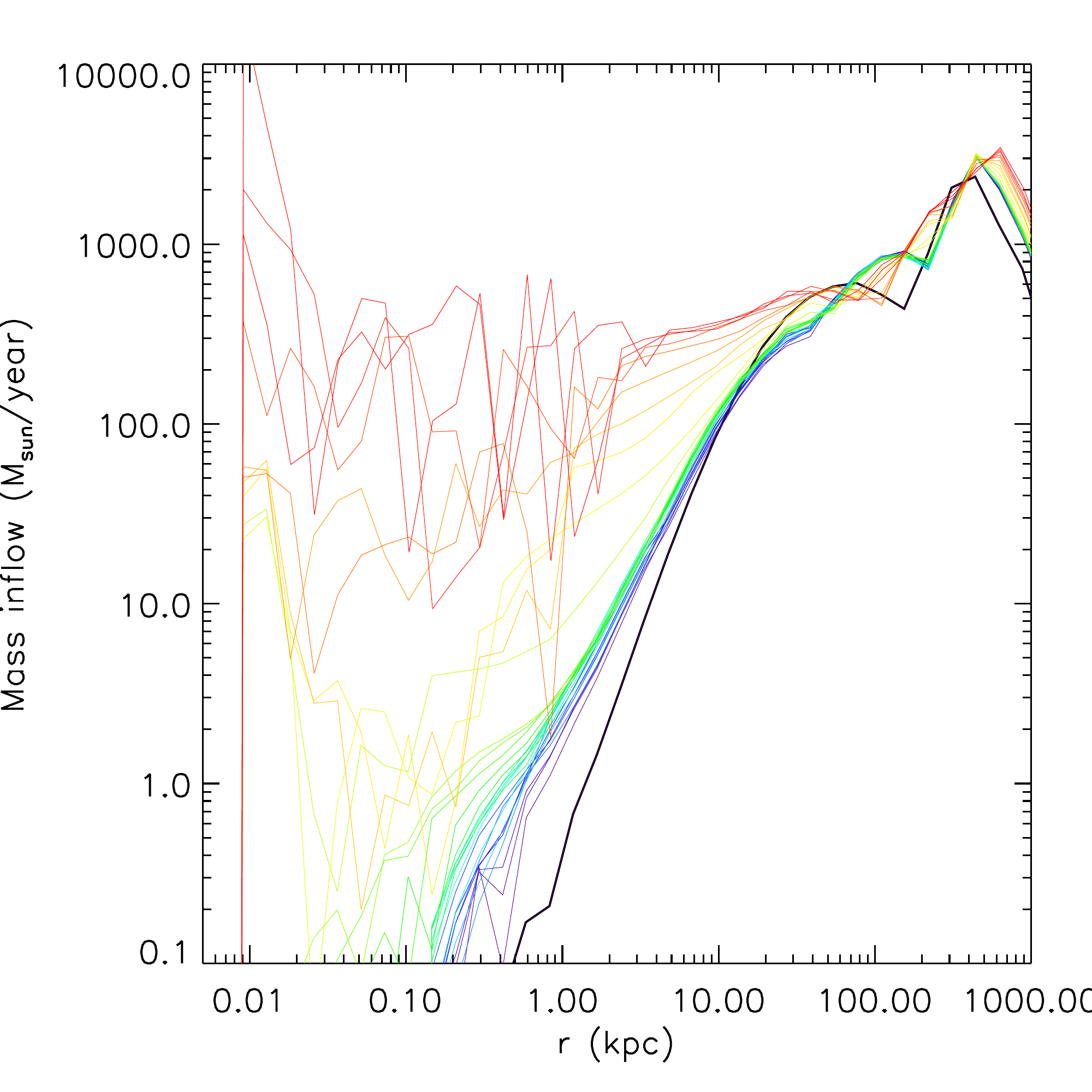}\includegraphics[scale=.4]{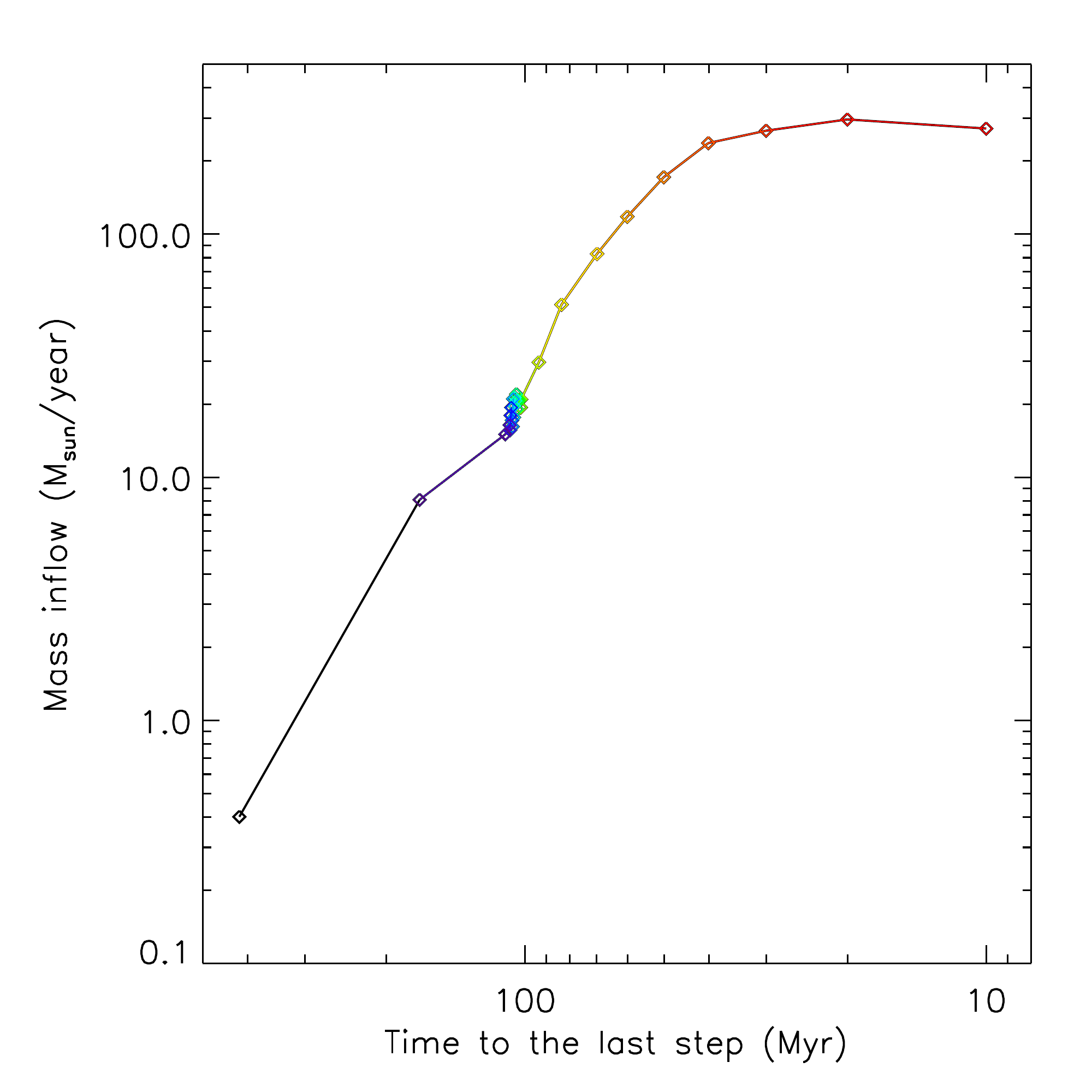}
\caption{The gas inflow of a simulation that goes $\sim 100$ Myr further than our standard run, reaching a steady state consistent with the classic cooling flow picture. The left panel shows the gas inflow profile at each output time and the right panel shows the time evolution of the inflow at a radius of 3.42 kpc. The radius is chosen to be large enough to avoid the fluctuation in the very center, while still providing a good estimate of the mass flow (see Section \ref{sec:comparison}). \label{fig_classic}}
\end{center}
\end{figure*}

In this section, we try to place these results in context, first making a link to steady-state cooling flow solutions, and then comparing to other (primarily simulation) work which looked specifically at only the developing cooling flow, and did not include feedback.

The classic cooling flow model \citep{Fabian94} predicts a ``cooling flow'' of $100s$ M$_{\odot}/$yr for rich clusters, assuming that in a steady state, without other heating sources, the gas flows inwards at a constant rate to replace the central gas that has cooled down and formed stars.  The mass drop-out occurs over the central cooling-flow region, and was originally assumed to cool and condense out in small clumps via a local cooling instability.   This picture has been known to be in disagreement with observations which usually indicate a star formation rate at least an order of magnitude lower than predicted by the steady state cooling flow model \citep{Tam01, ODea08, Rafferty08}.  Our simulations show that a steady state is not reached before the AGN feedback is potentially strong enough to balance cooling, and therefore we argue that this solution is not relevant.  However, it is interesting to see if we recover the steady-state result if we run the simulation for long enough. In Figure~\ref{fig_classic} we plot the gas inflow for a simulation with $N_{\rm root} = 64$ that runs much further than our major runs. We find that after less than a hundred Myr, without any heating mechanism, the system approaches a steady state with roughly constant mass flow of $\dot{M} \sim 300$ M$_{\odot}/$yr in the cluster core ($r < 100$ kpc), consistent with the classic cooling flow prediction. 

Most previous simulation work on cool core clusters has focused on the heating process, especially AGN feedback, but they do usually include a pure cooling flow simulation in which heating is turned off \citep[e.g.][]{Croton}. Our results are consistent with these results inasmuch as there is overlap.  For example, \citet{BM06} examined two dimensional models which also found that a cooling-only model results in a relatively flat temperature profile (falling only by a factor of 2-3 over a range of 100 in radius).  However, these simulations did not have the resolution ($\sim 1$ kpc) to follow the cooling catastrophe in detail, and instead used a parameterized mass drop-out term in the mass-conservation equation.  Similar results were found for a one-dimensional cooling flow in \citet{MB03}.

Our results are also consistent with previous theoretical work, for example, \citet{Bertschinger89} presented one-dimensional steady accretion models normalized to self-similar cooling waves that demonstrated slowly declining -- or even increasing -- temperature profiles.  They even found a sonic radius (similar in radius to our transition radius) which occurred very close to the SMBH.

Simulations which do not include a specific drop-out term and did not have high-resolution resulted in a much larger cooling region (a larger transition radius in our terminology) and a larger cooling flow immediately after the cooling catastrophe than found in our simulations. 

\citet{Kaiser03} provides a semi-analytic model of the time evolution of the cooling flow assuming hydrostatic equilibrium at every step of the evolution. Their density and temperature evolution agrees well with what we find for $r > 1$ kpc, when the inflow is subsonic and the gas is roughly in hydrostatic equilibrium. However, a semi-analytic model cannot describe what happens inside $r \lesssim 100$ pc where cooling runs away.

\citet{Guo08} analyses the local and global instabilities in CC clusters and shows that AGN feedback can suppress global radial instabilities. An interesting set of recent simulations \citep{McCourt11, Sharma11} have addressed the role of local cooling instabilities in simulations of a cool core, suppressing the cooling catastrophe by instituting a global heating term.  They showed that local cooling instabilities can form as long as the timescale for a cooling instability is less than 10 times the dynamical time.  That result does not conflict with our finding in this paper that local cooling instabilities do not occur before the global instability (or catastrophe) because of the global heating term instituted in those paper. 

Finally, we note that we do not include feedback and so do not compare to simulations which try to explicitly model AGN energy injection \citep[e.g.,][]{Omma04, Bruggen09, Dubois2010, Fabjan2010}, although we will examine this in future work.  Also, we have focused on the evolution of cool core clusters rather than the formation of cool core clusters, and so it remains possible that isotropic thermal conduction can play a role in the earlier evolution of a non-cool core clusters \citep{Voit11}.

% --------------------------------------------

\section{Conclusion}\label{sec:conclusion}

We performed high resolution simulations of an idealized, cool-core X-ray cluster loosely based on Perseus, in order to better understand how gas cools and accretes on to the SMBH.  The use of an AMR code was essential in order to resolve the important scales, ranging from the Mpc virial radius of the cluster, down to the pc scale, which is the size of the cold gas clumps.  We do not include feedback from the AGN and so the current work is not a complete description of cool-core cluster evolution; however, it does shed light on where cold gas condenses out of the flow, where heating needs to occur, and what observational signatures we would predict for such an object.  

We simulate an idealized cluster, but include small-scale velocity perturbations, as well as some net rotation, as implied by cosmological simulations.  We include radiative cooling, but no heating term.  The simulation resolves regions of high density and short cooling time.   We present our primary results below.

\begin{itemize}
 
\item We find that as the cluster core cools, the temperature profile remains remarkably flat, from a few hundred pc out to at least 20 kpc, with a change of only a factor of $\sim 2$ in temperature over this range.  We argue that this occurs because the gas flow adjusts such that adiabatic compression balances cooling at each radius over this range.  The gas that does cool does so via a global cooling catastrophe that occurs first at a `transition' radius of about 10 pc from the SMBH in our best resolved simulation.  This occurs about 300 Myr after the start of the simulation, which is about the cooling time of the gas in the initial configuration.

\item We stop the simulation about 20 Myr after the cooling catastrophe when the accretion rate onto the black hole (which we measure at a radius of 400 pc) becomes sufficiently large that a reasonable feedback efficiency could balance radiative cooling in the flow.  At this point, when we examine the amount of cold gas, there is a distinct lack of gas below a few keV (because of the flat temperature profile mentioned above), and is in agreement with observational constraints on Perseus.

\item  Although the simulation is intrinsically three-dimensional, the solution outside of the transition radius is well-described by spherically symmetric flow.  In particular, we find that local thermal instabilities do not grow in the cool-core cluster within 300 Myr (i.e. before the global cooling catastrophe), at least outside of the transition radius around $\sim 100$ pc. This is despite the fact that we seed the initial flow with substantial perturbations.

\item Isotropic thermal heat conduction does not significantly affect this result.  It does not lead to a large delay, nor does it significantly affect the density and temperature profile in the center.  Heating from Type Ia SN is similarly unimportant.

\item The final result is not very sensitive to the gas initial conditions, but is sensitive to the presence of the BCG.  Without a BCG, the gas temperature drops much more rapidly with radius, in conflict with observations.

\item We find that some aspects of the solution are quite sensitive to the resolution of the simulation.  In particular, we find that very high mass resolution is required in the gas near the cooling flow, and that lower resolution runs resulted in both a steeper slope in the temperature profile in the intermediate range from the transition radius to 20 kpc, as well a larger initial transition radius.

\end{itemize}

These results are intriguing for a number of reasons.  First, they imply that the real ``cooling-flow" problem arises very close to the center of the cluster (i.e. within a few hundred pc), in the sense that it is only in this region that gas cools to temperatures below those observed.  Therefore, the lack of cooler gas seen in X-ray spectroscopy may arise not because of the way that heating operates, but because of the temperature plateau caused (we argue) by the BCG potential well.   Mass drop-out occurs only at the very center of the cluster.

The fact that mass drop-out occurs first very close to the SMBH also provides a mechanism for thermal regulation to operate.   One of the unanswered questions for AGN regulation is how the properties of the gas in the entire cooling region, which stretches out to 100 kpc, can control what happens in the center hundred pc.  These simulations help to answer this by showing that this naturally occurs -- the lack of local thermal instabilities insures that the flow is coherent, and, as we show, the result is cold gas condensing out only inside the transition radius, where it can almost immediately form an accretion disk and feed the SMBH.

Of course, this doesn't relieve us of the need for energetic feedback, as we show that the mass inflow rate quickly grows and after a few hundred Myr, reaches hundreds of solar masses per year, in agreement with steady-state models, and in disagreement with observations.  We argue that, before this occurs, the accretion disk that we see feeds gas to the SMBH, generating `radio-mode' feedback that heats up the gas.  How exactly this feedback occurs is not clear and not addressed in this paper.  We will examine this point in more detail in future work.  Another aspect which is not explained by these results is the generation of H$\alpha$ filaments seen in most cool core clusters.  Here we simply note that they do not naturally occur at large radii via a local cooling instability before the flow becomes globally unstable.

\acknowledgments

We thank A.J.R. Sanderson for providing the observational data in Figure~\ref{fig_M_T}.  We also thank Mark Voit and Brian O'Shea for useful discussions, as well as the referee for suggestions which improved the presentation of this paper.  We acknowledge support from NSF grants AST-0547823, AST-0908390, and AST-1008134, as well as computational resources from NASA, the NSF Teragrid, and Columbia University's Hotfoot cluster.

% Figures

\end{document}